\documentclass[twocolumn,showpacs,amsfonts,aps,prc,floatfix]{revtex4} 

\usepackage{bm}
\usepackage{graphicx}
\usepackage{amsmath}
\usepackage{dcolumn}
\usepackage{wrapfig}

\def\La{\langle}
\def\Ra{\rangle}
\newcommand{\eq}{{\,=\,}}

\begin{document}

\title{Interplay of shear and bulk viscosity in generating
flow in heavy-ion collisions}
\date{\today}

\author{Huichao Song}
\email[Correspond to\ ]{song@mps.ohio-state.edu}
\affiliation{Department of Physics, The Ohio State University, 
Columbus, OH 43210, USA}
\author{Ulrich Heinz}
\affiliation{Department of Physics, The Ohio State University, 
Columbus, OH 43210, USA}

\begin{abstract}
We perform viscous hydrodynamic calculations in 2+1 dimensions to 
investigate the influence of bulk viscosity on the viscous suppression 
of elliptic flow in non-central heavy-ion collisions at RHIC energies.
Bulk and shear viscous effects on the evolution of radial and elliptic 
flow are studied with different model assumptions for the transport
coefficients. We find that the temperature dependence of the relaxation 
time for the bulk viscous pressure, especially its critical slowing down 
near the quark-hadron phase transition at $T_c$, partially offsets effects 
from the strong growth of the bulk viscosity itself near $T_c$, and that
even small values of the specific shear viscosity $\eta/s$ of the 
fireball matter can be extracted without large uncertainties from 
poorly controlled bulk viscous effects.
\end{abstract}

\maketitle

\section{Introduction}
\label{sec1}

A question of widespread interest is that of the specific shear
viscosity (shear viscosity per entropy density $\eta/s$) of the
quark-gluon plasma (QGP) created in nuclear collisions at the 
Relativistic Heavy Ion Collider (RHIC). Ideal ({\em i.e.} inviscid)
fluid dynamics has been quite successful in describing the transverse
momentum ($p_T$) spectra and the elliptic flow coefficient $v_2(p_T)$ 
of the bulk of the thousands of particles created in each, say, Au+Au 
collision \cite{Rev-hydro}. The agreement between theory and experiments 
improves further when one interfaces a (3+1)-dimensional ideal fluid 
description of the QGP phase with a hadron cascade during the late 
expansion stage, in order to properly account for the highly viscous 
evolution after hadronization of the QGP \cite{Teaney:2001av}. This 
success strongly suggests that the QGP fireball created at RHIC 
thermalizes very quickly and behaves like an almost perfect liquid 
\cite{Heinz:2001xi}, which implies that it must be a strongly coupled 
plasma \cite{Gyulassy:2004vg,Gyulassy:2004zy,Shuryak:2004cy}.

On the other hand, the quantum mechanical uncertainty relation places
a fundamental lower bound on the specific shear viscosity of any
medium \cite{Danielewicz:1984ww}, and explicit computation in a large
class of very strongly coupled quantum field theories (unfortunately
not including QCD) suggests that this limit is close to the so-called
KSS bound $\left.\frac{\eta}{s}\right|_\mathrm{KSS}=\frac{1}{4\pi}
\approx 0.08$ \cite{Policastro:2001yc,Kovtun:2004de}. While this
is a very small number (almost two orders of magnitude smaller than
that of any other known (real) fluid \cite{Kovtun:2004de,Csernai:2006zz},
with the possible exception of strongly interacting systems of ultracold 
fermionic atoms near the unitarity limit \cite{Schafer:2007pr,JETDUKE}), 
it is known that the anisotropic elliptic flow generated in non-central
relativistic heavy-ion collisions is very sensitive to shear viscosity
\cite{Heinz:2002rs,Teaney:2003kp}. The roots of this sensitivity lie
in the exceedingly rapid expansion of the heavy-ion collision fireballs,
especially during the early expansion stage which is characterized by 
very large components of the velocity shear tensor \cite{Danielewicz:1984ww}.
Recent progress in performing causal relativistic hydrodynamical simulations
of viscous fluids in 2+1 dimensions \cite{Heinz:2005bw,Romatschke:2007mq,%
Song:2007fn,Dusling:2007gi,Song:2007ux,Song:2008si,Luzum:2008cw,%
Molnar:2008xj,Molnar:2009tx,Teaney:2009zz,Heinz:2009xj,Romatschke:2009im,%
Teaney:2009qa} revealed that even very small specific shear viscosities,
close to the KSS bound, should leave easily identifiable experimental
signatures, in particular through a suppression of elliptic flow. That
in the experimental data this suppression in not large enough to lead
to immediate failure of the ideal fluid approach suggests that the QGP
viscosity, in the temperature region probed by RHIC collisions, must also
be close to the KSS bound \cite{Lacey:2006pn,Lacey:2006bc,Drescher:2007cd,%
Xu:2007jv,Xu:2008av}. 

Viscous hydrodynamics, in comparison with experimental data, allows in 
principle for an accurate determination of $\eta/s$. In practice, this 
requires excellent control of several other inputs that either are 
presently not known with sufficient accuracy or have not yet been 
correctly implemented in the numerical simulations \cite{Song:2008hj}.
The largest prevailing uncertainty is related to the initial source 
deformation that drives the elliptic flow which is presently not known
to better than 20-30\% \cite{Hirano:2005xf,Luzum:2008cw,Drescher:2006pi,%
Drescher:2006ca} (see, however, recent suggestions to eliminate this error 
source \cite{Lacey:2009xx,Heinz:2009cv}). As shown in \cite{Luzum:2008cw},
this leads at present to an ${\cal O}(100\%)$ uncertainty in the extracted
$\eta/s$ value. Two other effects of similar magnitude which, however,
work against each other and may largely cancel, are strong viscous
effects \cite{Hirano:2005xf} and the non-equilibrium chemical composition
\cite{Kolb:2002ve,Teaney:2002aj,Hirano:2002ds,Hirano:2005wx,Huovinen:2007xh} 
in the late hadronic phase. Finally (and this is the point of the present 
paper) bulk viscous effects must be included in any study that aims to 
extract the specific shear viscosity \cite{Song:2008hj,Song:2009je}. However,
even when making generous allowances for all these uncertainties, it
appears clear that the effective shear viscosity to entropy ratio of the
QGP, averaged over the expansion history of the fireballs created in RHIC 
collisions, cannot exceed the conservative upper limit 
\begin{eqnarray*}
 \left.\frac{\eta}{s}\right|_\mathrm{QGP} < 5\times \frac{1}{4 \pi}.
\end{eqnarray*}
This makes the QGP the most perfect liquid ever observed in the 
laboratory.

In the present paper we use the (2+1)-dimensional viscous relativistic 
fluid dynamic code {\tt VISH2+1} \cite{Song:2007fn,Song:2007ux,Song:2008si}
to study the effects of bulk viscosity and their interplay with shear 
viscosity in the buildup of radial and elliptic flow. Some preliminary
results were reported in \cite{Song:2008hj,Song:2009je} (see also 
\cite{Denicol:2009am} for related work). Our work is preceded by three 
(0+1)-dimensional studies for systems undergoing boost-invariant 
longitudinal expansion without transverse flow 
\cite{Torrieri:2008ip,Fries:2008ts,Rajagopal:2009yw} which explored the 
suggestion by Torrieri, Tom\'a\v{s}ik and Mishustin \cite{Torrieri:2007fb} 
that in rapidly expanding fireballs bulk viscosity can lead to such large 
negative bulk pressures that the fluid becomes mechanically unstable
against clustering and cavitation. Since bulk viscosity is expected to
be maximal near the quark-hadron phase transition (see discussion in
Section~\ref{sec2}), those studies predicted that bulk viscous effects
become important mostly during the second half of the fireball expansion
when the QGP undergoes hadronization. At that time the scalar expansion 
rate $\theta{\,\equiv\,}\partial_\mu u^\mu$ (where $u^\mu(x)$ the flow 
4-velocity), which for 1-dimensional boost-invariant expansion equals 
$\theta\eq{1}/\tau$ (where $\tau\eq\sqrt{t^2{-}z^2}$ is the longitudinal 
proper time, with $z$ indicating the longitudinal or beam direction), is 
already small enough to significantly temper the growth (in magnitude) of 
the (negative) bulk pressure, leading to instability problems only for 
relatively large peak values for the specific bulk viscosity $\zeta/s$ 
\cite{Torrieri:2008ip,Fries:2008ts,Rajagopal:2009yw}. 

Our work improves on these analysis by including a realistic initial 
transverse density profile and the resulting transverse flow in the 
fireball. This has two important consequences: (i) The transverse flow 
increases the expansion rate $\theta$, leading to larger bulk pressures 
$|\Pi|$ for given $\zeta/s$. (ii) Some of the matter near the dilute 
transverse edge of the fireball experiences large bulk viscosities 
already at very early times where the expansion rate $\theta\sim{1}/\tau$ 
is big. This leads to much more severe problems with mechanical instability 
in {\tt VISH2+1} than for simple 1-dimensional expansion, and to 
correspondingly smaller values for the upper limit for $\zeta/s$ that 
allows for stable hydrodynamic evolution. Even more restrictive than the 
condition for mechanical stability is the self-consistency constraint for 
the validity of viscous hydrodynamics itself: the entire framework, which 
is based on a near-equilibrium expansion, breaks down when viscous 
corrections to the local equilibrium distribution function become 
comparable to the thermal equilibrium terms. We will show that this 
happens, for particles with typical momenta $p{\,\sim\,}3T$, even 
before the effective total pressure becomes negative and mechanical 
instability sets in \cite{fn1}. While the formalism may be able to 
qualitatively indicate where and when cavitation sets in 
\cite{Torrieri:2008ip,Fries:2008ts,Rajagopal:2009yw}, we doubt that the 
phenomenon itself can be self-consistently described within the existing 
viscous hydrodynamic frameworks.

Obviously, viscous hydrodynamics can predict the viscous suppression of 
elliptic flow reliably only within its domain of validity. We therefore 
restrict our attention to the parameter range where the bulk viscous 
pressure stays everywhere sufficiently small that stable hydrodynamic 
evolution is ensured. Within that range (which we determine), we study 
the effects of bulk viscosity and of the microscopic relaxation time for 
the bulk viscous pressure on radial and elliptic flow, with and without 
additional shear viscosity. For a fluid with constant specific shear 
viscosity $\frac{\eta}{s}\eq\frac{1}{4\pi}$ we find that, depending on 
initial conditions and details of the temperature dependence of the 
relaxation time, bulk viscosity increases the viscous suppression of 
$v_2(p_T)$ by 5 -- 50\%. This large range indicates not only that bulk 
viscosity is a potentially serious contaminant in the extraction of the 
specific shear viscosity $\eta/s$ from elliptic flow data, but also that 
a robust theoretical effort is needed to better constrain the range of 
reasonable values for the bulk viscosity and its associated relaxation 
time. We find that the uncertainty range is drastically reduced, to the 
$10-20\%$ level, if we impose proportionality between the specific bulk 
viscosity and its associated relaxation time, as indicated by kinetic 
theory. The critical growth of the specific bulk viscosity near the 
quark-hadron phase transition is then accompanied by critical slowing 
down of the dynamics of the viscous bulk pressure. This diminishes the 
bulk viscous contribution to the viscous suppression of elliptic flow.    

The paper is organized as follows: In Section~\ref{sec2} we review the 
present state of knowledge of the temperature dependence of the specific 
bulk and shear viscosities $\zeta/s$ and $\eta/s$ and their associated 
microscopic relaxation times. Based on this analysis we introduce 
specific parametrizations for $(\zeta/s)(T)$ and the bulk pressure 
relaxation time $\tau_{_\Pi}(T)$ which we use later in the numerical 
simulations. Section~\ref{sec3} gives a brief summary of specific 
features of the viscous hydrodynamic equations solved in this work, 
referring to earlier work for a more general description. In 
Section~\ref{sec4} we discuss generic effects of bulk and shear 
viscosity on the hydrodynamic evolution of fireball eccentricity 
and flow and their implications for the final hadron spectra and 
elliptic flow. Section~\ref{sec5} discusses the sensitivity of bulk 
viscous effects to the initialization of the bulk viscous pressure
and to its relaxation time. In Section~\ref{sec6} we explore the range
of bulk viscosities that allows for stable viscous hydrodynamic evolution.
Consequences of bulk viscous effects for the extraction of the specific
shear viscosity $\eta/s$ from experimental elliptic flow data are
discussed in Section~\ref{sec7} before summarizing our findings in 
Section~\ref{sec8}.   

\section{Viscosities and relaxation times}
\label{sec2}

The present state of knowledge of the viscous properties of strongly
interacting matter at high temperatures is nicely reviewed in 
\cite{Kapusta:2008vb} to which we refer for details. Kinetic theory 
\cite{non-rel} and experiment \cite{Kovtun:2004de,Csernai:2006zz} show 
that for non-relativistic fluids the {\sl specific shear viscosity}
$\eta/s$ typically reaches a minimum near the liquid-gas phase transition,
rising both towards lower temperatures in the liquid phase and towards
higher temperatures in the gas phase. Lattice QCD \cite{Meyer:2007ic}, 
perturbative QCD \cite{Arnold:2003zc}, and hadron cascade simulations 
\cite{Demir:2008tr} indicate that relativistic QCD matter behaves
analogously, but with the liquid and gas phases interchanged (the 
liquid QGP phase exists at higher temperature than the hadronic gas
phase). According to perturbative \cite{Arnold:2003zc} and lattice 
\cite{Meyer:2007ic} QCD, the increase of $\eta/s$ with temperature in 
the QGP phase is weak over the temperature range explored in RHIC 
collisions, suggestion the use of a constant $\eta/s$ for the QGP in 
hydrodynamic simulations. We here use the smallest value for this 
constant permitted by the KSS conjecture \cite{Kovtun:2004de},
$\left.\frac{\eta}{s}\right|_\mathrm{KSS}=\frac{1}{4\pi} \approx 0.08$, 
in order to extract reasonable upper bounds for the uncertainties
introduced by bulk viscous effects into the extraction of such a small 
value from experimental data. [We note that the assumption of a constant
$\eta/s$ is unacceptable for quantitative attempts to extract it
from heavy-ion collision data; at least one must account for a
significant increase of this ratio during hadronization and in the late
hadronic phase \cite{Demir:2008tr}.]

The {\sl relaxation time $\tau_\pi$ for the shear viscous pressure tensor}
$\pi^{\mu\nu}$ has been computed for a relativistic Boltzmann gas 
\cite{Israel:1976tn,Baier:2006um}, in weakly coupled QCD \cite{York:2008rr}, 
in lattice QCD \cite{Meyer:2009jp}, and in ${\cal N}\eq4$ SYM theory at 
infinite coupling \cite{Baier:2007ix,Bhattacharyya:2008jc,Natsuume:2007ty}. 
The results can be presented in the form $\tau_\pi\eq\frac{\lambda}{T}
\frac{\eta}{s}$, with $\lambda$ bracketed by 
$2{\,\lesssim\,}\lambda{\,\lesssim\,}6$. We here use
$\tau_\pi = \frac{3}{T}\frac{\eta}{s} = \frac{1.5}{2\pi T}$.

Theoretical knowledge of the {\sl specific bulk viscosity $\zeta/s$} is 
more murky. For a non-interacting system of massless quanta it vanishes
exactly, due to conformal invariance. Interactions lead to deviations
from zero that usually remain small, except near phase transitions where 
the system may develop large correlation lengths \cite{Onuki,Paech:2006st,%
Moore:2008ws}. Kinetic theory gives $\frac{\zeta}{\eta}\eq\kappa 
\left(\frac{1}{3}{-}c_s^2\right)^2$ where $\kappa\eq5/3$ in relaxation time 
approximation \cite{Gavin:1985ph} and $\kappa\eq15$ for a system 
of photons radiated by massive particles in thermal equilibrium 
\cite{Weinberg:1971mx}. The complete leading order result for weakly 
coupled QCD \cite{Arnold:2006fz} is roughly consistent with the latter 
of these two, but adds a weak (decreasing) temperature dependence. At 
high temperatures $c_s^2{\,\approx\,}\frac{1}{3}$, so the ratio
$\frac{\zeta}{\eta}$ is small of second order in the deviation. For 
strongly coupled ${\cal N}\eq4$ SYM theory one obtains a lower bound 
for this ratio which is only linear in this deviation and thus much 
larger: $\frac{\zeta}{\eta}{\,\geq\,}2\left(\frac{1}{3}{-}c_s^2\right)$
\cite{Buchel:2007mf}. For the hadron gas different authors 
\cite{Prakash:1993bt,Davesne:1995ms,Chen:2007kx} agree that 
$\frac{\zeta}{\eta}{\,\ll\,}1$ just below $T_c$ and that this ratio 
decreases towards lower temperatures. There is no agreement on the sign
of the temperature dependence of the specific bulk viscosity $\zeta/s$
itself which according to \cite{Prakash:1993bt} increases towards lower 
temperature for massive pions, but decreaes for massless pions
\cite{Chen:2007kx}. However, there are general arguments 
\cite{Paech:2006st,Kharzeev:2007wb} that support the idea that $\zeta/s$
should peak near the quark-hadron phase transition, due to long-range
correlations related to the restoration of chiral symmetry; at a
second-order critical point $\zeta/s$ is predicted to diverge 
\cite{Moore:2008ws,Buchel:2009hv}.

%
\begin{figure}[htp]
\includegraphics[width=\linewidth,clip=]{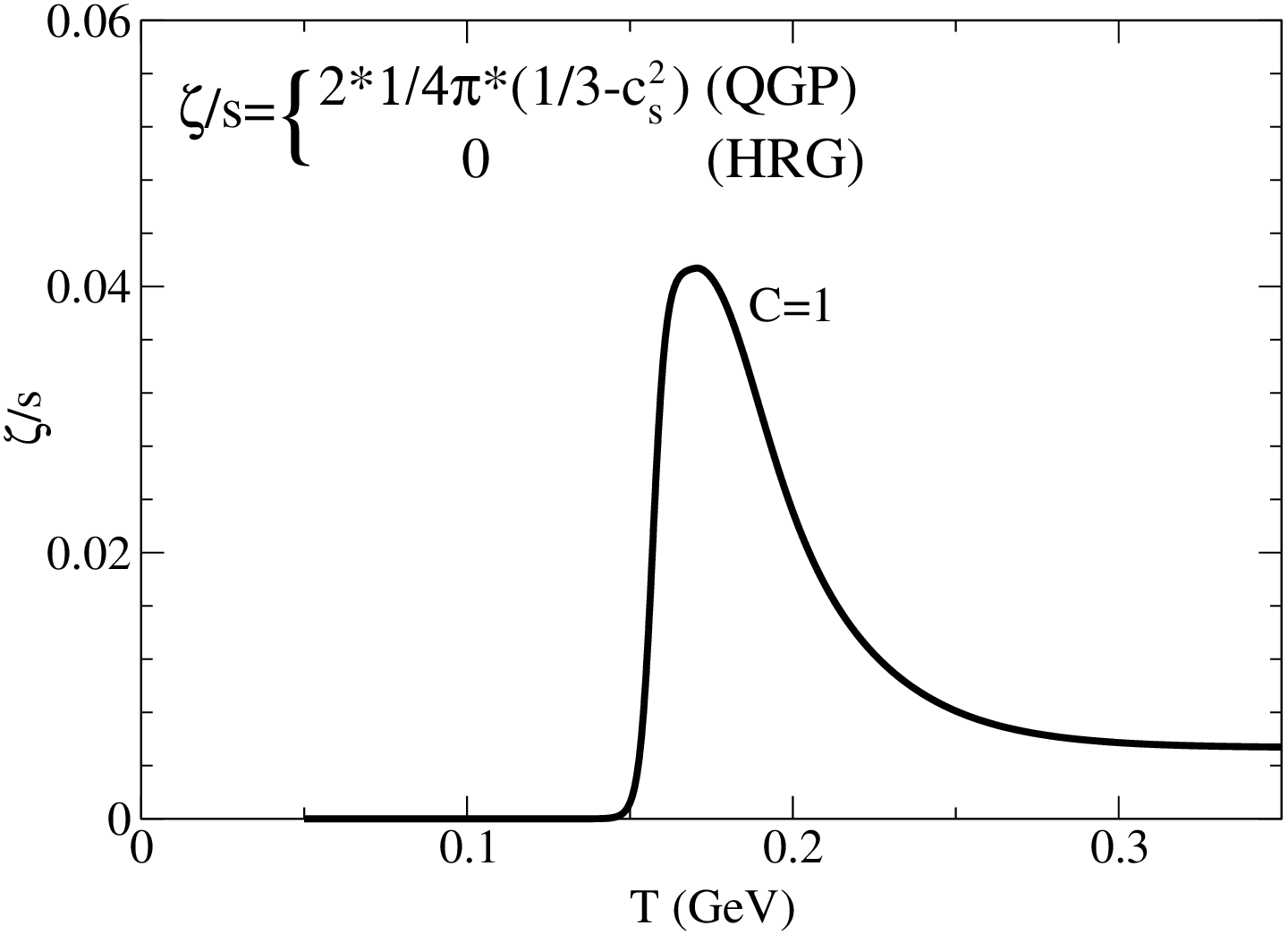}
\caption{\label{F1}
(Color online) Parametrization of the specific bulk viscosity 
$\zeta/s$ as a function of temperature. ($C$ is a multiplicative scaling 
factor for the entire function, see text.)
}
\vspace*{-5mm}
\end{figure}
%

We assume here that $\zeta/s$ quickly approaches zero once $T$ decreases 
below $T_c$; above $T_c$, we parametrize it as 
$\frac{\zeta}{s}\eq\frac{1}{2\pi}\left(\frac{1}{3}-c_s^2\right)$
(which corresponds to the Buchel bound \cite{Buchel:2007mf} for 
$\eta/s=1/(4\pi)$), using $c_s^2(T)$ extracted from the same lattice 
QCD data \cite{Katz:2005br} that we used for our equation of state 
(EOS~L, see \cite{Song:2008si} for details). The factor 
$\left(\frac{1}{3}-c_s^2\right)$ increases as we approach $T_c$ from
above; the resulting increase of $\zeta/s$ is qualitatively, but not
quantitatively consistent with a direct extraction of $\zeta/s$ from
lattice QCD \cite{Meyer:2007dy} (for a critical discussion of this
extraction see \cite{Moore:2008ws}) and with recent work in ``holographic
QCD'' \cite{Gursoy:2009kk}. We connect our parametrization above $T_c$ 
to the assumed zero value for $\zeta/s$ well below $T_c$ by interpolating 
with a Gaussian function (see Fig.~\ref{F1}).
This results in a peak value $(\zeta/s)(T_c)\simeq0.04$ -- about half
as big as our choice for the (temperature independent) specific shear 
viscosity $\eta/s$ and consistent with strong coupling estimates for
strongly coupled conformal field theories using the AdS/CFT correspondence 
\cite{Gubser:2008yx} (which was also used by Buchel when deriving his bound)
and with holographic QCD \cite{Gursoy:2009kk}. It is, however, more than 
10 times smaller than both the lattice QCD value extracted by Meyer 
\cite{Meyer:2007dy} and a recent AdS/CFT-based estimate by Buchel 
for a non-conformal plasma \cite{Buchel:2009hv}. We will see that this
factor 10 has crucial implications for the applicability of viscous 
hydrodynamics. To simulate larger bulk viscosities, we scale the 
function shown in Fig.~\ref{F1} (to which we will refer as ``minimal 
bulk viscosity'' for brevity) by a constant factor $C{\,>\,}1$.

Finally, we must specify the {\sl relaxation time $\tau_{_\Pi}$ for the bulk
viscous pressure} $\Pi$ about which even less is known theoretically.
In Israel-Stewart theory (both in its macroscopic and microscopic kinetic 
formulation \cite{Israel:1976tn}) one has $\tau_{_\Pi}\eq\zeta\beta_0$
where $\beta_0$ is some combination of thermal equilibrium integrals.
This suggests that, if $\zeta/s$ peaks near $T_c$ due to growing
correlation lengths, so does the relaxation time $\tau_{_\Pi}$ for the bulk 
pressure (``critical slowing down'' \cite{Dusling-priv}). Buchel 
\cite{Buchel:2009hv} found that in theories where the specific heat 
diverges at $T_c$, $c_V{\,\sim\,}1/\sqrt{|1{-}T_c/T|}$, the relaxation
time can actually diverge at $T_c$ even if $\zeta/s$ remains finite,
{\it i.e.} as $T{\,\to\,}T_c$, $\tau_{_\Pi}$ grows more strongly than 
$\zeta/s$ (see also the discussion in \cite{Moore:2008ws}). We use the 
parametrization 
\begin{equation}
\label{eq1}
  \tau_{_\Pi}(T) = \max\left[\tilde\tau{\cdot}\frac{\zeta}{s}(T),
                          \,0.1\,\mathrm{fm}\right],
  \ \text{with}\ 
  \tilde\tau = 120\,\text{fm}/c.
\end{equation}
This increases linearly with $\zeta/s$ as $T{\,\to\,}T_c$, but imposes 
a non-zero lower bound on $\tau_{_\Pi}$, for reasons of numerical stability
of {\tt VISH2+1}. For comparison we also study two constant relaxation 
time values, $\tau_{_\Pi}\eq0.5$ and 5\,fm/$c$, roughly corresponding to 
the smallest and largest values of Eq.~(\ref{eq1}) for temperatures
$1{\,\leq\,}T/T_c{\,\leq\,}2$ if we set $C\eq1$.
  
\section{Viscous hydrodynamics}
\label{sec3}

We solve the following second order viscous hydrodynamic equations
(``Israel-Stewart equations'' \cite{Israel:1976tn,Muronga:2001zk,%
Muronga:2004sf,Muronga:2006zw}),
\begin{eqnarray}
\label{eq2}
&&\!\!\!\!\!\!\!\! 
d_\mu T^{\mu\nu}=0,
\ \ 
T^{\mu\nu} = e u^\mu u^\nu - (p{+}\Pi)\Delta^{\mu\nu} + \pi^{\mu\nu},
\\
\label{eq3}
&&\!\!\!\!\!\!\!\!
\Delta^{\mu\alpha}\Delta^{\nu\beta} D\pi_{\alpha\beta}
=-\frac{1}{\tau_{\pi}}(\pi^{\mu\nu}{-}2\eta\sigma^{\mu\nu})
\nonumber\\
&&\!\!\!\!\!\!\!\!
\phantom{\Delta^{\mu\alpha}\Delta^{\nu\beta} D\pi_{\alpha\beta}=}
 -\frac{1}{2}\pi^{\mu\nu} \frac{\eta T}{\tau_\pi}
       d_\lambda\left(\frac{\tau_\pi}{\eta T}u^\lambda\right),
\\ 
\label{eq4}
&&\!\!\!\!\!\!\!\!
D \Pi
=-\frac{1}{\tau_{\Pi}}(\Pi+\zeta \theta)
 -\frac{1}{2}\Pi\frac{\zeta T}{\tau_{_\Pi}}
       d_\lambda\left(\frac{\tau_{_\Pi}}{\zeta T}u^\lambda\right),
\end{eqnarray}
in the two transverse spatial directions and time ((2+1)-d), implementing
boost-invariant longitudinal expansion along the beam direction. We 
assume zero net baryon density and thus vanishing heat conductivity.
Here, $T^{\mu\nu}$ is the energy momentum tensor, $\pi^{\mu\nu}$ 
is the shear pressure tensor, and $\Pi$ is bulk pressure. $d_\mu$
denotes the covariant derivative components (see 
\cite{Heinz:2005bw,Molnar:2009tx} for details) in the curvilear coordinates 
$(\tau, x, y, \eta_s)$ where $\tau=\sqrt{t^2{-}z^2}$ is the longitudinal
proper time and $\eta_s=\frac{1}{2}\ln\frac{t{+}z}{t{-}z}$ is the
space-time rapidity. $\Delta^{\mu\nu} =g^{\mu\nu}{-}u^\mu u^\nu$
projects onto the spatial components in the local rest frame (here
$g^{\mu\nu}=\text{diag}(1,\,-1,\,-1,\,-1/\tau^2)$ is the metric tensor
in $(\tau, x, y, \eta_s)$ coordinates); $\nabla^{\mu}\eq\Delta^{\mu\nu}
d_\nu$ is the spatial gradient and $D\eq u^\mu d_\mu$ is the time 
derivative in that frame. The driving forces for the shear and bulk 
viscous pressures are the (symmetric and traceless) velocity stress tensor 
$\sigma^{\mu\nu}\eq\nabla^{\left\langle\mu\right.} u^{\left.\nu\right
\rangle}{\,\equiv\,}\frac{1}{2}(\nabla^\mu u^\nu{+}\nabla^\nu u^\mu)
-\frac{1}{3}\Delta^{\mu\nu}\theta$ and the scalar expansion rate 
$\theta\eq{d}_\nu u^\nu\eq\nabla_\nu u^\nu$, respectively. The shear and
bulk viscosities $\eta$ and $\zeta$ and their associated relaxation times
$\tau_{\pi}$ and $\tau_{\Pi}$ were discussed in the preceding Section.

The last terms in Eqs.~(\ref{eq3}) and (\ref{eq4}) are of second order
in deviations from local equilibrium. For conformal systems they can be
written in various equivalent forms, up to higher order corrections
\cite{Song:2008si}. Even for non-conformal systems, such as QCD with
the equation of state EOS~L used here (see below), the difference between
the terms as written down here and their various conformal approximations 
are numerically insignificant \cite{Song:2008si} unless inconsistently
large relaxation times are used. Other second order terms that should
be allowed for on the right hand sides of Eqs.~(\ref{eq3}) and (\ref{eq4}) 
were identified in \cite{Baier:2007ix,Betz:2009zz}, and some of their
coefficients were derived in the weak coupling limit in \cite{York:2008rr}. 
Recent code verification efforts by the TECHQM Collaboration \cite{TECHQM}
indicate that these additional terms have very little numerical influence.
We therefore ignore them in the present study.

The explicit form of Eqs.~(\ref{eq2}-\ref{eq4}) for longitudinally boost 
invariant ({\it i.e.} $\eta_s$-independent) systems is given in 
\cite{Heinz:2005bw}. The equations are closed by providing an equation
of state for which we use EOS~L as described in Ref.~\cite{Song:2008si}.
We study Au+Au collisions with the same initial conditions for the 
starting time $\tau_0\eq0.6$\,fm/$c$ and Glauber model initial energy 
density profiles as in Ref.~\cite{Song:2008si}, with peak density 
$e_0{\,\equiv\,}e(r{=}0,\tau_0;b{=}0)\eq30$\,GeV/fm$^3$ in central 
($b{=}0$) collisions. For the viscous pressures we use either zero 
($\Pi(x,y,\tau_0;b)\eq\pi^{\mu\nu}(x,y,\tau_0;b)\eq0$) or
Navier-Stokes initial conditions ($\Pi(x,y,\tau_0;b)\eq{-}\zeta
\theta(\tau_0)$ and $\pi^{\mu\nu}(x,y,\tau_0;b)\eq2\eta 
\sigma^{\mu\nu}(\tau_0)$), calculated from the initial velocity profile 
(which does not depend on $x,\,y$ and $b$, due to the absence of initial 
transverse flow). The actual choice will be noted when discussing the 
results. As in \cite{Song:2008si} we end the hydrodynamic evolution and
compute the final hadron spectra on a freeze-out surface of constant
temperature $T_\mathrm{dec}\eq130$\,MeV.

\section{Viscous evolution and spectra: generic features}
\label{sec4}

In this section we compare generic effects on the hydrodynamic 
evolution and final particle spectra caused by shear and bulk viscous 
effects separately. (Their combined effects will be explored in 
Sects.~\ref{sec5}-\ref{sec7}.) To this end we perform hydrodynamic 
comparison runs for central ($b\eq0$) and non-central ($b\eq7$\,fm) 
Au+Au collisions, using identical initial and freeze-out conditions, 
for (i) an ideal fluid, (ii) a viscous fluid with only minimal shear 
viscosity $\frac{\eta}{s}\eq\frac{1}{4\pi}$, and (iii) a viscous fluid
with only ``minimal bulk viscosity'' ($C\eq1$) as defined in 
Sec.~\ref{sec2}. In the viscous runs we choose Navier-Stokes initial 
conditions for the viscous pressures and equal relaxation times 
$\tau_\pi\eq\tau_{_\Pi}\eq\frac{3\eta}{sT}$. (As a caveat we note that 
in the bulk viscous case results can depend sensitively on the initial 
conditions, depending on the characteristics of the relaxation time 
for the bulk viscous pressure -- see discussion in Sec.~\ref{sec5}.)
The results for case (ii) supplement those for the smaller Cu+Cu collision 
system studied in \cite{Song:2007fn,Song:2007ux} (although for a more 
realistic equation of state and using Eq.~(\ref{eq3}) instead of the 
``simplified Israel-Stewart equation'' employed in those earlier papers).

\subsection{Hydrodynamic evolution}
\label{sec4a}

\begin{figure}[t]
\includegraphics[width=\linewidth,clip=]{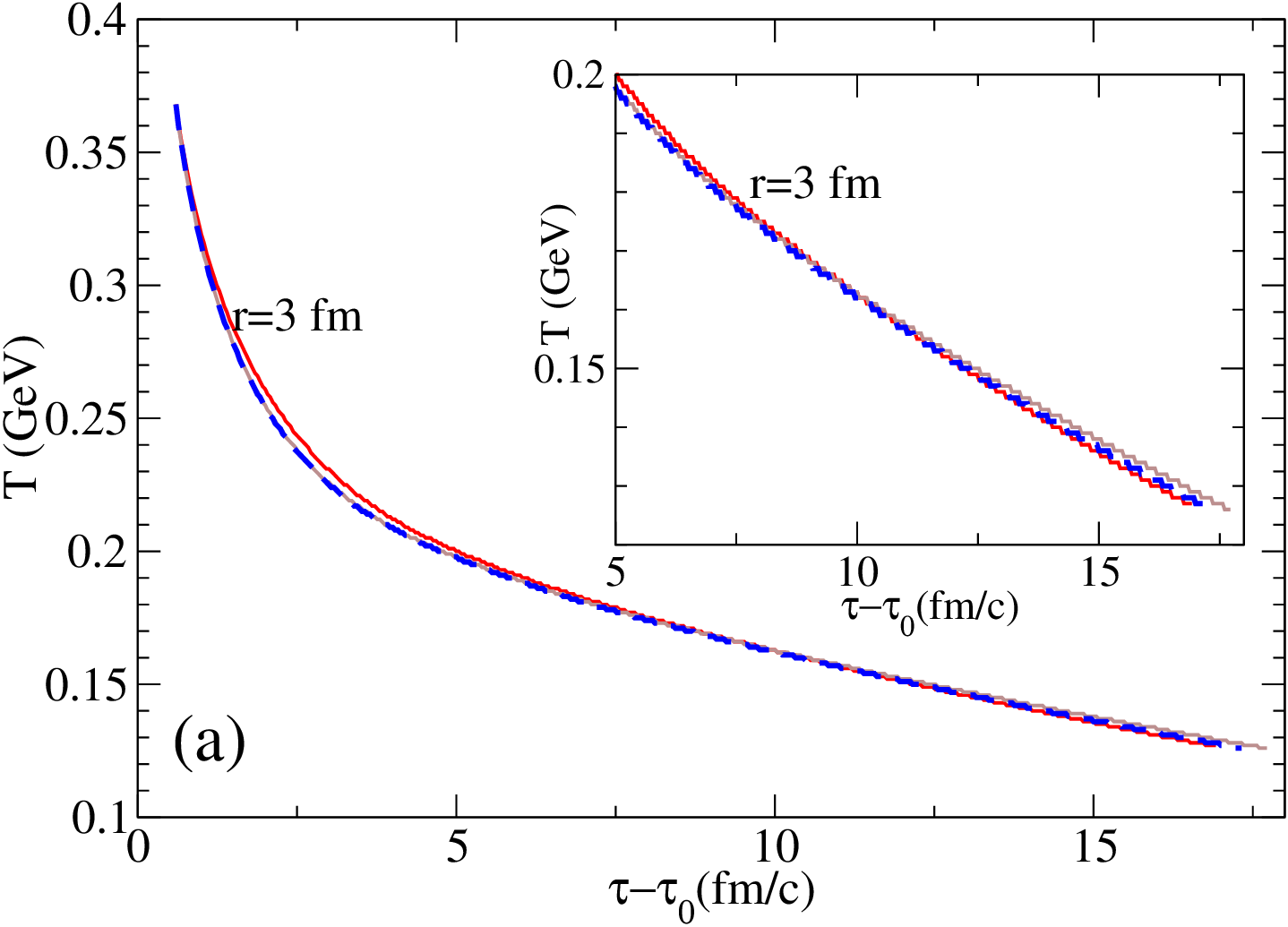}\\
\includegraphics[bb=19 33 706 522,width=\linewidth,clip=]{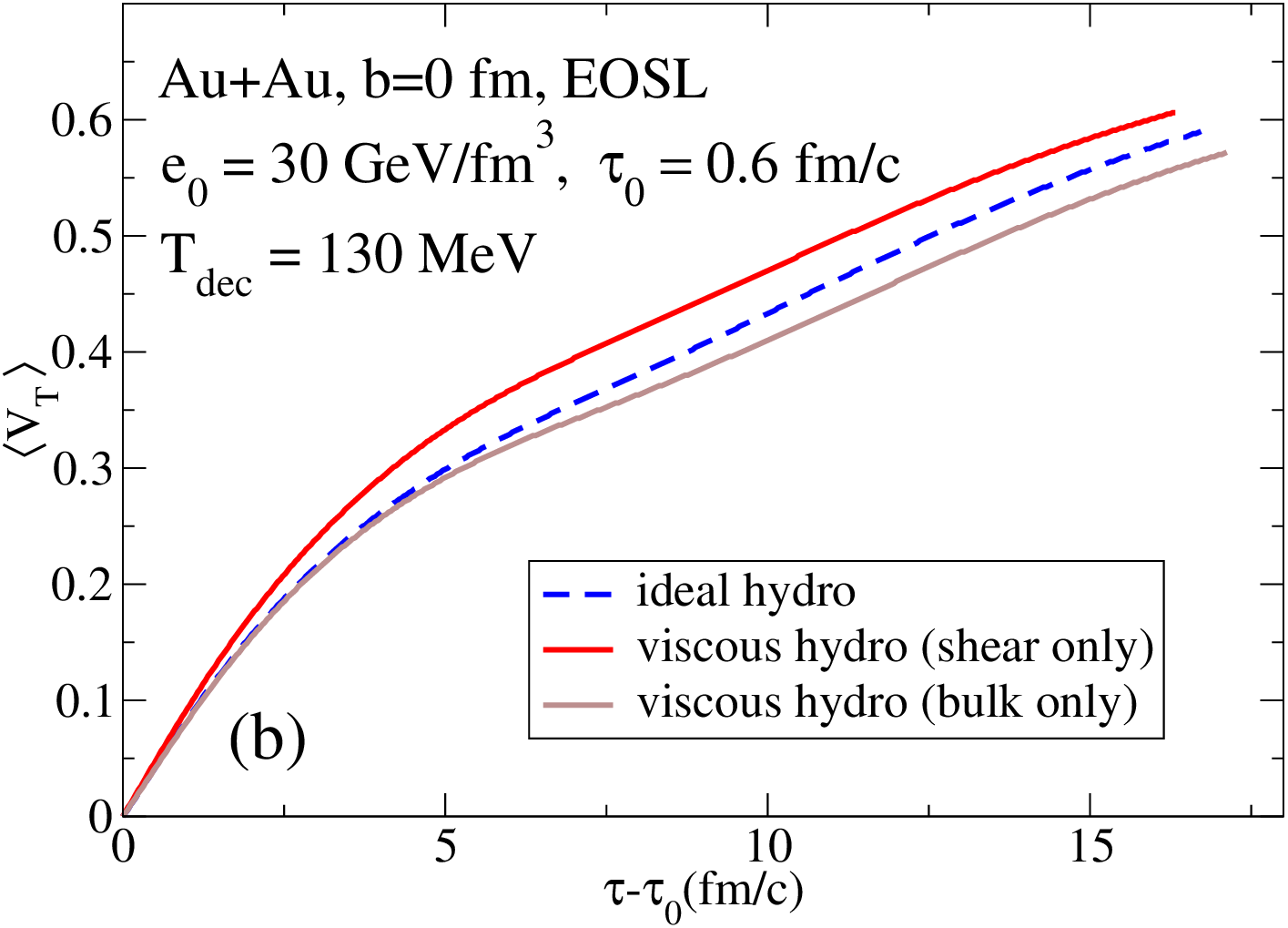}
\caption{\label{F2} (Color online) 
Time evolution of the local temperature (a) and average radial 
flow (b), from ideal hydrodynamics (dashed blue), viscous hydrodynamics with 
only minimal shear (solid red) or bulk (solid brown) viscosity. In (b) the 
average radial flow is calculated with the Lorentz contracted energy 
density $\gamma_\perp e$ in the transverse plane as weight function. The 
inset in (a) shows the late evolution with increased resolution.
}
\end{figure}

Figure~\ref{F2}a shows the time evolution of the local temperature
in central Au+Au collisions for the three cases. (We plot the temperature 
at a radius $r\eq3$\,fm from the fireball center since at $r\eq0$ the 
curves for cases (i) and (iii) are almost indistinguishable.) Compared 
with the ideal fluid, shear viscosity reduces the work done by longitudinal 
pressure and thus slows down the cooling process during the early stage; 
during the middle and late stages, shear viscosity accelerates the cooling
since the positive transverse shear pressure leads to stronger radial flow
than for the ideal fluid (see Fig.~\ref{F2}b and Ref.~\cite{Song:2007ux}
for a full discussion). At late times, the shear viscous fireball thus 
cools more rapidly than the ideal fluid \cite{Song:2007fn,Song:2007ux}.

Bulk viscosity, on the other hand, reduces the work done in all three
directions, due to the isotropic negative bulk pressure 
$\Pi{\,\sim\,}{-}\zeta\theta$ driven by the positive expansion rate 
$\theta{\,>\,}0$. As a result, radial flow develops less rapidly than 
for the ideal fluid (Fig.~\ref{F2}b), and the bulk viscous fluid cools 
(slightly) more slowly than the ideal one during all expansion stages 
(Fig.~\ref{F2}a). While the expansion rate is largest at very early times, 
the bulk viscosity then is very small throughout the fireball, except for 
a thin region near the transverse edge of the fireball where the matter 
is close to $T_c$; bulk viscous effects are therefore almost negligible
until most of the matter cools down to $T_c$. At this time the longitudinal
expansion rate has significantly decreased, but transverse expansion 
picks up some of the the slack, and we see significant bulk viscous 
effects on radial flow evolution between 5 and 10\,fm/$c$. Surprisingly, 
the consequences for the cooling rate are significantly smaller than in 
the shear viscous case: with the parameters studied here, the cooling 
rates for the ideal and bulk viscous fluids almost agree. 

We now turn to non-central collisions. To describe the fireball
deformations in configuration and momentum space, we use the spatial
eccentricity $\varepsilon_x\eq\frac{\langle\!\langle y^2{-}x^2\rangle\!
\rangle}{\langle\!\langle y^2{+}x^2\rangle\!\rangle}$ (where 
$\langle\!\langle \dots\rangle\!\rangle$ denotes an energy density
weighted average over the transverse plane \cite{Kolb:1999it}) and the 
momentum anisotropies $\varepsilon_p\eq\frac{\langle T_0^{xx}{-}T_0^{yy}
\rangle}{\langle T_0^{xx}{+}T_0^{yy}\rangle}$ (defined in terms of 
unweighted averages over the transverse plane of components of the ideal 
fluid part of the energy-momentum tensor, and thus measuring only the 
collective flow anisotropy \cite{Song:2007ux}) and 
$\varepsilon_p'\eq\frac{\langle T^{xx}{-}T^{yy}\rangle}
{\langle T^{xx}{+}T^{yy}\rangle}$ (defined in terms of the total 
energy-momentum tensor that contains the viscous pressure components 
and thus additionally includes microscopic momentum anisotropies in the 
local rest frame of the fluid \cite{Song:2007ux}).

%
\begin{figure}[t]
\includegraphics[width=\linewidth,clip=]{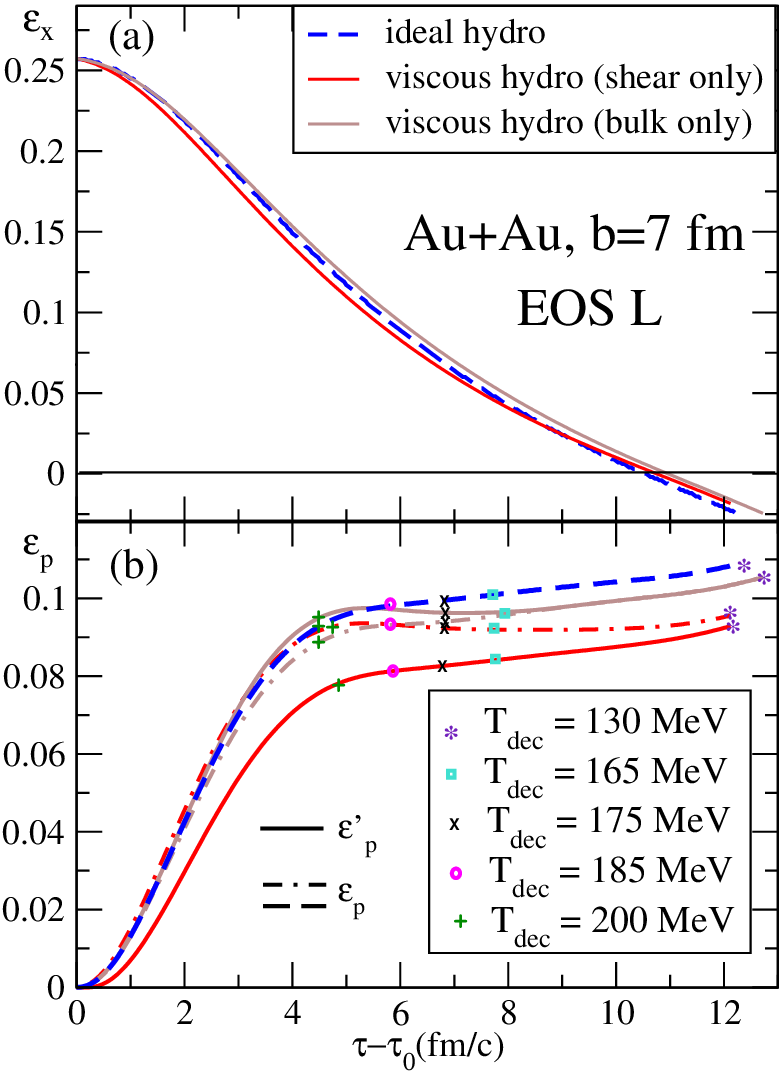}
\caption{\label{F3} (Color online)
Time evolution of spatial eccentricity $\varepsilon_x$ (a) and momentum 
anisotropy $\varepsilon_p$, $\varepsilon_p'$ (b), from ideal and viscous 
hydrodynamics (see text for details). In (b), the different symbols 
along the bulk viscous fluid lines indicate central ($r{=}0$) freeze-out 
times for different freeze-out temperatures as described in the legend.
} 
\end{figure}
%

\begin{figure*}[bth]
\vspace*{-2mm} 
\includegraphics[width=0.49\linewidth,clip=]{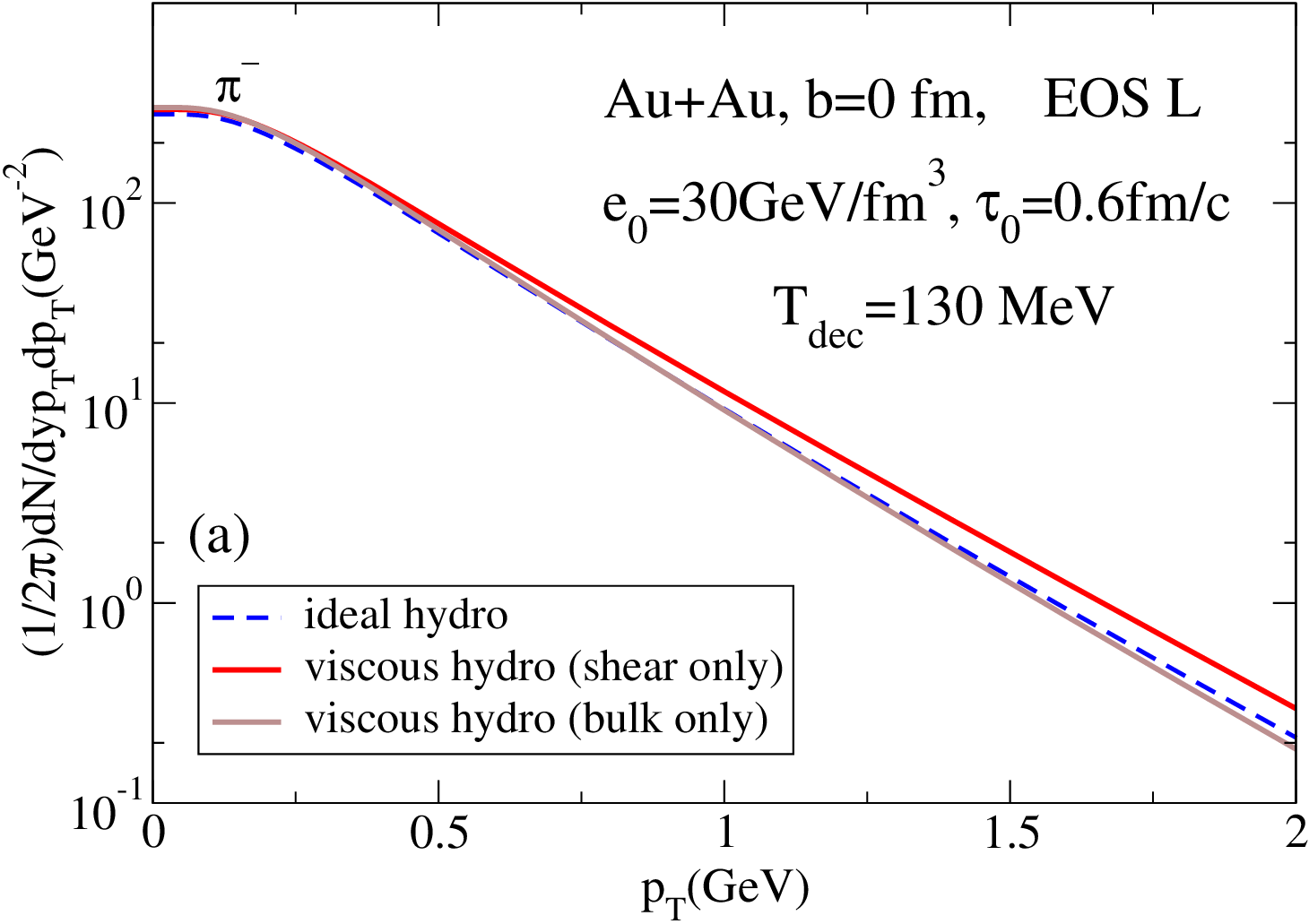}
\includegraphics[width=0.48\linewidth,clip=]{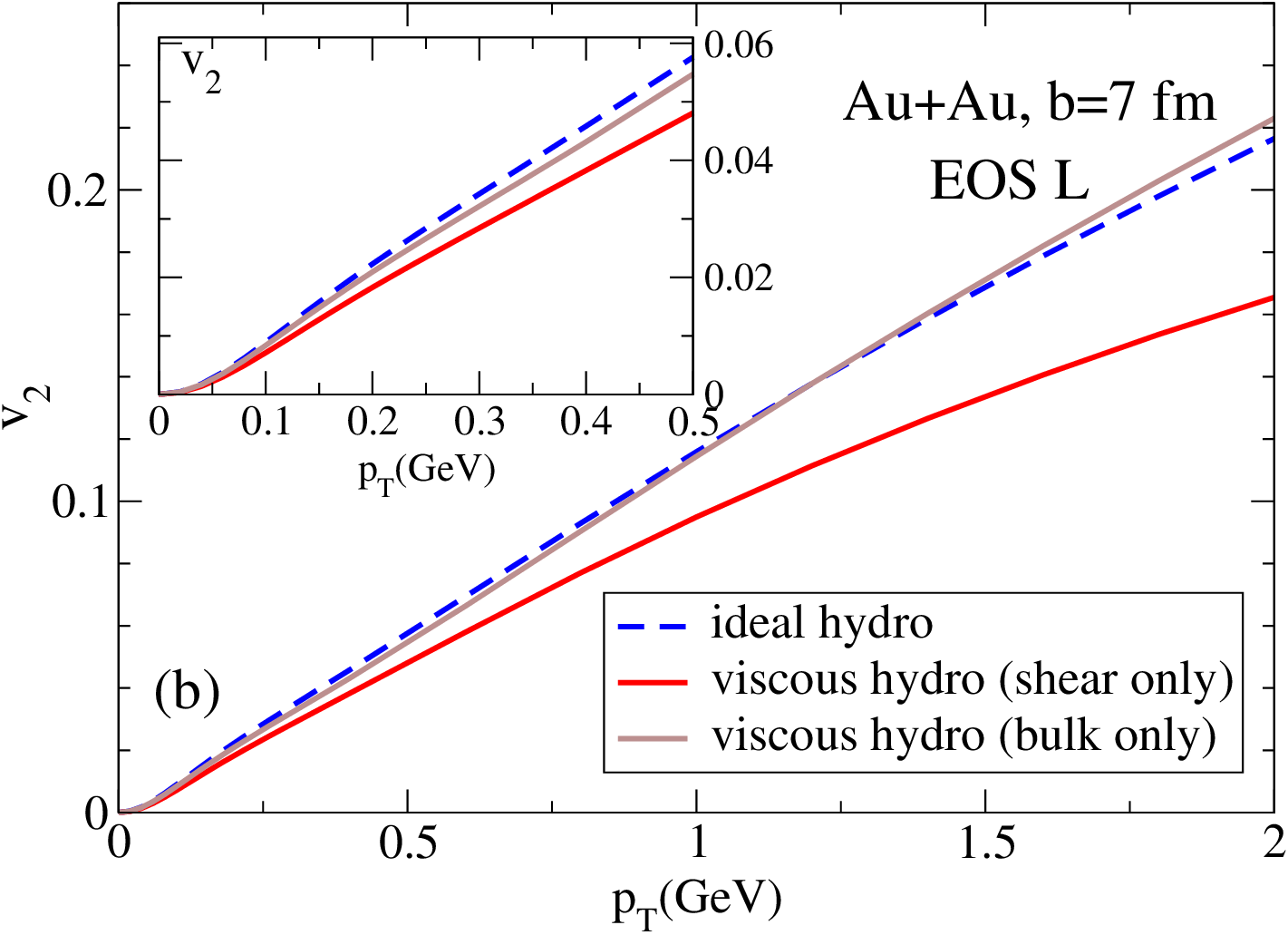}
\caption{\label{F4} (Color online) 
$p_T$ spectra and elliptic flow $v_2(p_T)$ for directly emitted pions 
({\it i.e.} without resonance decay contributions).
}
\vspace*{-3mm} 
\end{figure*}

Figure~\ref{F3}a shows the time evolution of the spatial eccentricity 
$\varepsilon_x$ for non-central Au+Au collisions at $b\eq7$\,fm. Compared 
with the ideal fluid, bulk viscosity decelerates whereas shear viscosity
initially accelerates the decrease with time of the spatial eccentricity 
$\varepsilon_x$. This is a direct consequence of the weaker radial flow 
in the bulk viscous case and the stronger radial flow in the shear viscous 
case. It is easy to see that {\em isotropic} radial expansion is enough to 
decreases the spatial eccentricity $\varepsilon_x$ \cite{Kolb:1999it}; 
{\em anisotropic} flow, with larger flow velocities in the reaction plane 
than perpendicular to it, only accelerates the rate with which it decreases. 
At late times, the eccentricity for the shear viscous fluid decreases more 
slowly than for the ideal one, since the ideal fluid develops {\em stronger 
elliptic flow} (see Fig.~\ref{F3}b and following discussion). In contrast, 
in the bulk viscous case the slower rate of decrease of the eccentricity 
is caused by {\em weaker radial flow}.

As the spatial eccentricity of the fireball decreases, its momentum
anisotropy increases. This is shown in Fig.~\ref{F3}b. The dash-dotted
lines for $\varepsilon_p$, which takes only the ideal fluid part
$T_0^{\mu\nu}\eq(e{+}p)u^\mu u^\nu-pg^{\mu\nu}$ of the energy-momentum
tensor into account, show how hydrodynamic forces convert the spatial 
anisotropy into a flow anisotropy. We observe that at early times the
flow anisotropy $\varepsilon_p$ rides on the developing radial flow:
compared to the ideal fluid, $\varepsilon_p$ develops a little faster
in the shear viscous fluid but a little more slowly in the bulk viscous 
case, following the evolution of radial flow. 

The difference between $\varepsilon_p$ (dash-dotted lines) and 
$\varepsilon_p'$ (solid lines) stems from the viscous pressure 
components in the energy-momentum tensor. It reflects a contribution 
to the momentum anisotropy that does not arise from anisotropic 
collective flow but from viscous deviations from isotropy of the 
microscopic momentum distribution $f\eq{f}_\mathrm{eq}{+}\delta f$
in the local fluid rest frame \cite{Song:2007ux}, accounted for by the 
non-ideal terms in $T^{\mu\nu}$. Figure~\ref{F3}b shows that for the 
shear viscous fluid these viscous corrections are large and {\em 
negative} (reflecting local momentum anisotropies pointing {\em out 
of the reaction plane} \cite{Song:2007ux}, {\it i.e.} opposite to the 
flow anisotropy), especially at early times when the expansion rate and 
shear velocity components $\sigma^{\mu\nu}$ are large. In contrast, the 
bulk viscous fluid shows significant viscous corrections only after about 
2.5\,fm/$c$, lasting until about 8\,fm/$c$, which is when in these 
non-central ($b\eq7$\,fm) Au+Au collisions the bulk of the matter passes 
through the phase transition where $\zeta/s$ is large ({\it c.f.} the 
temperature markers on the curves shown in Fig.~\ref{F3}b). The bulk 
viscous pressure contribution to $\varepsilon_p'$ is {\em positive}, 
{\it i.e.} pointing {\em into} the reaction plane, parallel to the 
collective flow anisotropy. (This was also recently pointed out by 
Monnai and Hirano \cite{Monnai:2009ad}.) 

For both shear and bulk viscosity, the sign of the viscous pressure 
contributions to $\varepsilon_p'$ obeys a ``Lenz rule'': they act 
{\em against} the radial flow driven effects on the momentum anisotropy. 
At late times these contributions become small in both cases, in the bulk 
viscous case driven by the rapid disappearance of $\zeta$ below 
$T_c$ (as modeled by us), in the shear viscous case by the more gradual
vanishing of the shear velocity tensor $\sigma^{\mu\nu}$ \cite{Song:2007ux}. 
We see that in the bulk viscous case the radial flow effect on 
$\varepsilon_p'$ eventually wins over that from the local deviation
from equilibrium $\delta f$, whereas the opposite is true for the shear 
viscous fluid. In both cases, the net effect at freeze-out is thus a 
viscous {\em suppression} of the momentum anisotropy below the ideal 
fluid limit (see purple stars in Fig.~\ref{F3}b). The viscous suppression 
of $\varepsilon_p'$ resulting from shear viscosity is 4-5 times stronger 
that that arising from bulk viscosity.   

\subsection{Spectra and elliptic flow}
\label{sec 4b}

From the hydrodynamic output at decoupling temperature $T_\mathrm{dec}$ 
the spectra and their azimuthal anisotropy, in particular the elliptic 
flow coefficient $v_2(p_T)$, are computed with a modified Cooper-Frye 
algorithm that takes into account that in viscous hydrodynamics the 
local phase-space distribution  $f(x,p)$ on the freeze-out surface 
slightly deviates from thermal equilibrium, $f\eq{f}_\mathrm{eq}{+}\delta 
f$, due to small but non-zero viscous pressure components 
\cite{Teaney:2003kp,Romatschke:2007jx,Song:2007ux,fn2}. Figure~\ref{F4}a shows 
the pion $p_T$-spectra for central Au+Au collisions from ideal and viscous 
hydrodynamics. Compared to the spectrum from ideal fluid dynamics, shear 
viscosity leads to flatter spectra while bulk viscosity generates steeper 
ones. This is a direct reflection of the stronger radial flow caused by 
the positive transverse shear pressure and the weaker radial flow 
resulting from the negative bulk viscous pressure. The slightly larger 
normalization of the viscous spectra is a consequence of viscous entropy 
production which leads to larger final multiplicities 
\cite{Romatschke:2007jx,Song:2008si}.

Figure~\ref{F4}a shows that shear and bulk viscosity act against each 
other in how they affect the slope of the $p_T$-spectra. When both 
viscosities are included together in the viscous calculations, this 
reduces the amount of readjustment needed in the initial conditions 
when fitting the measured transverse momentum spectra with viscous 
instead of ideal fluid dynamics \cite{Romatschke:2007jx}. The differential
elliptic flow $v_2(p_T)$ for soft pions, on the other hand, is affected 
by both bulk and shear viscosity in the same way: Figure~\ref{F4}b
shows that the viscous reduction of $\varepsilon_p'$ in Fig.~\ref{F3}b
translates directly into reduced elliptic flow $v_2$ of the final
hadron spectra. This is true in particular in the low-$p_T$ region
$p_T{\,<\,}1$\,GeV/$c$ (see inset in Fig.~\ref{F4}b). At larger $p_T$,
the shear viscous $v_2$ suppression further increases, due to a negative 
contribution from $\delta f$ along the freeze-out surface. In contrast, 
bulk viscosity increases $v_2(p_T)$ above 1\,GeV/$c$ because it results 
in steeper $p_T$ spectra. At $p_T\eq0.5$\,GeV/$c$ (approximately the mean
transverse momentum for pions) we find that minimal bulk viscosity 
suppresses $v_2$ by ${\sim\,}5\%$ while minimal shear viscosity
leads to a ${\sim\,}20\%$ suppression. 

If this were the complete story, the additional ${\sim\,}5\%$ 
bulk viscous $v_2$ suppression would lead to a ${\sim\,}25\%$ 
reduction of the value for $\eta/s$ that one might extract from
experimental elliptic flow data, by comparing them with an ideal 
fluid dynamical baseline as advertised in \cite{Luzum:2008cw}.
This is a non-negligible effect. Since the input for the bulk 
viscosity used in our calculations is fraught with large theoretical
uncertainties, as discussed in Sec.~\ref{sec2}, this points to a 
likelihood for correspondingly large uncertainties in the empirical
extraction of $\eta/s$ from experiment. In the following Section we
follow this line of thought further, by investigating the additional
sensitivity of bulk viscous effects to the initial conditions for 
the bulk viscous pressure and to its relaxation time.  

\section{Sensitivity of bulk viscous dynamics to initial conditions 
and relaxation times}
\label{sec5}

In this Section we now focus entirely on bulk viscosity and investigate 
what happens when we change the initial value for the bulk viscous 
pressure $\Pi$ and its relaxation time $\tau_{_\Pi}$. We study only
``minimal bulk viscosity'' as defined in Sec.~\ref{sec2} ({\it i.e.}
$C\eq1$), leaving the discussion of larger values to the following 
Section.

%
\begin{figure}[t]
\vspace{-0.0cm}
\includegraphics[bb=35 25 724 523,width=\linewidth,clip=]{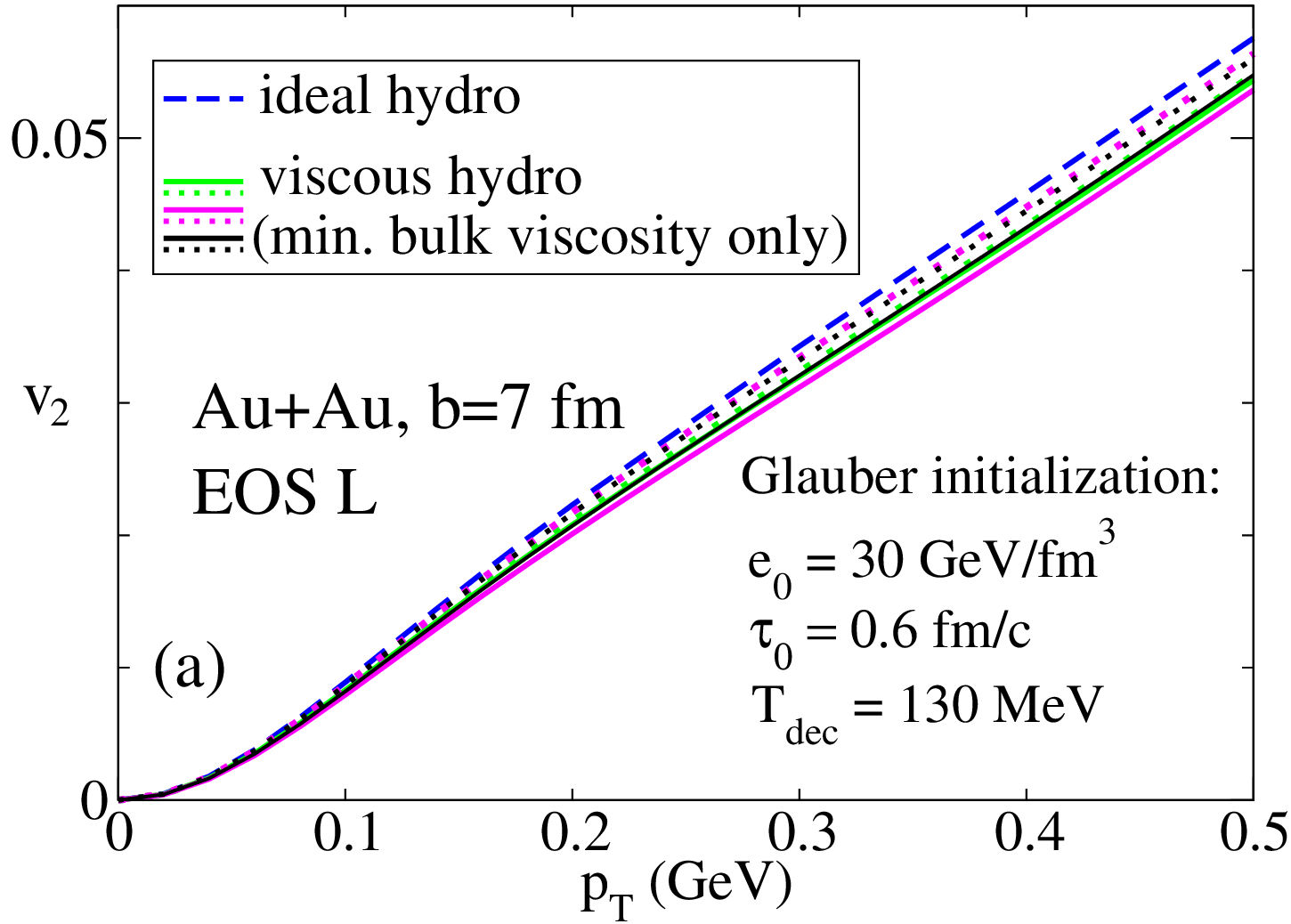} \\
\includegraphics[width=\linewidth,clip=]{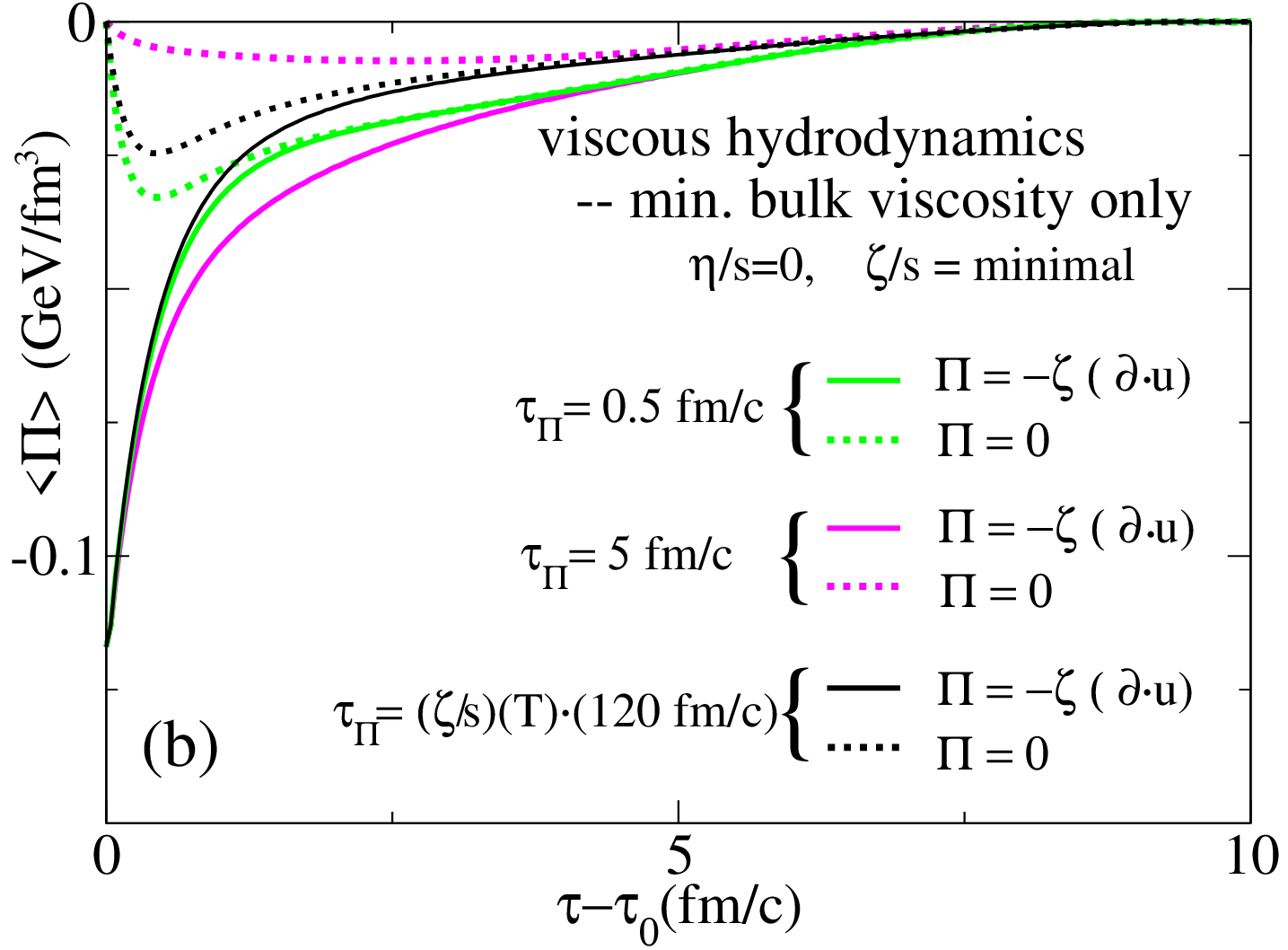}
\caption{\label{F5}
(Color online) (a) Differential elliptic flow $v_2(p_T)$ 
for directly emitted pions (without resonance decays) from ideal and 
viscous hydrodynamics, including only minimal ($C\eq1$) bulk viscosity. 
(b) Time evolution of the bulk pressure $\La\Pi\Ra$ averaged 
over the transverse plane (weighted by the energy density) from viscous 
hydrodynamics. Different curves correspond to different initializations 
and relaxation times, as indicated (see text for discussion).
} 
\end{figure}
%

Figure~\ref{F5} shows, for peripheral Au+Au collisions at $b\eq7$\,fm, the 
differential elliptic flow for pions (a) and the time evolution of the 
average value of the bulk viscous pressure $\La\Pi\Ra$ (b), for the two 
initial conditions (zero and Navier-Stokes) for $\Pi$ and the three 
choices of relaxation time scales $\tau_{_\Pi}$ discussed in Sec.~\ref{sec2}.
At $p_T\eq0.5$\,GeV/$c$, Fig.~\ref{F5}a indicates a bulk viscous $v_2$ 
suppression that ranges (for $C\eq1$) from ${\sim\,}2\%$ to ${\sim\,}10\%$.
For the short relaxation time $\tau_{_\Pi}\eq0.5$\,fm/$c$ (solid and
dotted green lines) the suppression is insensitive to the initialization 
of $\Pi$, yielding about 8\% suppression below the ideal fluid value 
at $p_T\eq0.5$\,GeV/$c$ for both zero and Navier-Stokes initial values.
Figure~\ref{F5}b explains the underlying reason for this observation:
for this short relaxation time, $\La\Pi\Ra$ quickly loses all memory of 
its initial value, relaxing for both initial conditions to the same 
trajectory after ${\sim\,}1-2$\,fm/$c$ ({\it i.e.} after a few 
relaxation times, similar to what we saw earlier \cite{Song:2007ux}
for the shear pressure components). This also demonstrates that most of 
the finally observed bulk viscous $v_2$ suppression is generated during 
the middle part of the expansion when most of the matter cools through 
the phase transition. If it were dominated by large negative bulk 
viscous pressures in the outer shell of the fireball at early times, 
we should see stronger sensitivity to the initial value for $\Pi$.

This changes completely if we chose a 10 times longer relaxation time,
$\tau_{_\Pi}\eq5$\,fm/$c$ (solid and dotted magenta lines in Fig.~\ref{F5}).
Now the bulk viscous $v_2$ suppression becomes extremely sensitive
to the initialization of $\Pi$. For zero initialization, the average 
bulk pressure $\La\Pi\Ra$ always remains small, leading to very small
(${\cal O}(2\%)$) final suppression effects for $v_2$. For Navier-Stokes
intialization, $\La\Pi\Ra$ is initially very large and negative (due to 
the large initial expansion rate) and, instead of relaxing to smaller
values as predicted by Navier-Stokes theory (and realized by the
solid green line corresponding to short $\tau_{_\Pi}$), it remains 
larger than the N-S value for about 4\,fm/$c$. As seen in Fig.~\ref{F5}a,
this leads to a much larger $v_2$ suppression of about 10\%.

The ``critical slowing down'' scenario which uses a temperature dependent
relaxation time that follows the behaviour of $\left(\frac{\zeta}{s}\right)
\!(T)$ is shown by the solid and dash-dotted black curves in Fig.~\ref{F5}. 
In this case the bulk viscous pressure quickly relaxes to its Navier-Stokes
value in the interior of the fireball where the temperature is high and
the relaxation time is short; near the edge of the fireball, however, where 
the temperature is near $T_c$ and the relaxation time is long, it remembers
its initial value (either zero or the large negative initial N-S value) for 
a long period. With some reflection one convinces oneself that this implies 
that for {\em both} zero and N-S initializations the magnitude of the average 
bulk viscous pressure $\La\Pi\Ra$ remains {\em below} the value observed for 
the short relaxation time. This is seen in Fig.~\ref{F5}b when comparing 
the black and green curves. Correspondingly, the viscous $v_2$ suppression 
seen in part (a) of the Figure is for both initializations smaller for the 
``critical slowing down'' scenario than for a short constant relaxation 
time. When comparing the ``critical slowing down'' scenario with the long 
constant relaxation time, the viscous $v_2$ suppression is significantly
smaller for N-S initialization (${\cal O}(7\%)$ vs. ${\cal O}(10\%)$) and 
about equally small (${\cal O}(2\%)$) for zero initialization.  

Since these findings contradict at least our own naive first expectations, 
we briefly reiterate the main point: taking into account the critical 
slowing down of the bulk viscous pressure dynamics near $T_c$ where 
$\zeta/s$ becomes large leads to {\em weaker} bulk viscous suppression 
effects on the elliptic flow than seen for both short and long {\em 
constant} ({\it i.e.} $T$-independent) relaxation times $\tau_{_\Pi}$. 

\section{Larger $\bm{\zeta/s}$ and the breakdown of viscous fluid 
dynamics}
\label{sec6}

As noted in Sec.~\ref{sec2}, the peak value of $\zeta/s$ in our 
parametrization shown in Fig.~\ref{F1} is about 10 times smaller than
some other estimates \cite{Meyer:2007dy,Buchel:2009hv}. When one tries
to simply multiply the function shown in Fig.~\ref{F1} by $C\eq10$, one
finds that (except for special circumstances discussed below) the
viscous hydrodynamic code crashes. The reason is that sufficiently large 
bulk viscosity can lead to fireball regions where the effective total 
isotropic pressure $p{+}\Pi$ (thermal + bulk viscous pressure) becomes
negative and the medium becomes mechanically unstable and will tend to 
break up \cite{Torrieri:2008ip,Fries:2008ts,Rajagopal:2009yw,Torrieri:2007fb}.
In fact, since certain components of the shear viscous pressure (in 
particular its longitudinal component $\pi^{\eta\eta}$) are usually 
also negative, instability can set in even somewhat earlier 
\cite{Fries:2008ts,Rajagopal:2009yw}. In numerical simulations this 
manifests itself through the exponential amplification of local 
numerical errors which will eventually stop the code from running. 

We point out that even before the fluid becomes mechanically unstable
one has left the region of applicability of viscous hydrodynamics. The
viscous hydrodynamic formalism is based on a near-equilibrium expansion;
its validity assumes that the viscous corrections to the energy-momentum
tensor are small compared with the ideal fluid terms. In other words,
if the condition $(|\Pi|{+}|\pi^{\mu\nu}|)/(e{+}p){\,\ll\,}1$ is 
violated for any component $(\mu\nu)$, the evolution based on equations
(\ref{eq2}-\ref{eq4}) can no longer be trusted. Ignoring the shear pressure
and setting $e{+}p\eq{sT}{\,\approx\,}4p$ for a QGP, the instability
threshold $p{+}\Pi\eq0$ translates into $|\Pi|/(e{+}p){\,\approx\,}\frac
{1}{4}$ which is not sufficiently small to trust the continued validity
of the equations. The following alternate consideration leads to the same
conclusion: If the fluid can be described by quasiparticles, the viscous
terms in the energy-momentum tensor correspond to deviations of the
local phase-space distribution $f(x,p)\eq{f}_\mathrm{eq}{+}\delta f$
from local equilibrium. Using Grad's 14-moment method, the deviation
$\delta f$ is expanded up to quadratic order in momentum 
\cite{Israel:1976tn,Teaney:2003kp,Baier:2006um,Betz:2009zz,Monnai:2009ad}
and (for a fluid with only bulk viscosity and massless particles at 
midrapidity $y\eq0$) can thus be written in the form 
\begin{equation}
\label{deltaf}
  \frac{\delta f}{f_\mathrm{eq}} = a\, \frac{p_T^2}{T^2}\, 
  \frac{\Pi}{e{+}p},
\end{equation}
where $a$ is a slowly varying function of temperature with magnitude 
of order unity \cite{Monnai:2009ad}. When $p{+}\Pi\eq0$ such that 
$\frac{\Pi}{e{+}p}\eq{-}\frac{1}{4}$, this means that for midrapidity
particles with typical thermal momenta $p_T{\,\simeq\,}3T$ the deviation 
$\delta f/f_\mathrm{eq}$ is negative with magnitude 1 or larger, rendering 
the total distribution function $f$ negative, which is unphysical. Clearly
the deviations from local equilibrium are too large and the formalism
breaks down.

In this Section we explore the range of bulk viscosities that are allowed 
without leaving the region of validity of second-order (Israel-Stewart) 
viscous hydrodynamics. As in the preceding Section, we study both zero 
and Navier-Stokes initial conditions and the same three choices for the 
bulk viscous relaxation time $\tau_{_\Pi}$, but we now also include runs 
where the fluid has an additional shear viscosity $\eta/s\eq(1\div2)/(4\pi)$,
with shear viscous relaxation time $\tau_\pi\eq3\eta/(sT)$, and we vary the
time $\tau_0$ when we start the hydrodynamic evolution. (For later starting 
times, we downscale the initial peak entropy density $s_0$ such that
the total initial entropy $\sim s_0 \tau_0$ is held constant.) For the 
specific bulk viscosity $\left(\frac{\zeta}{s}\right)\!(T)$ we take the 
functional form shown in Fig.~\ref{F1}, but multiplied by an arbitrary 
constant $C{\,>\,}1$. For each set of initial conditions, $\tau_{_\Pi}$, 
and $\eta/s$ we determine the largest value $C_\mathrm{max}$ that still 
allows for stable running of the code, {\it i.e.} where the effective 
total isotropic pressure $p{+}\Pi$ does not violate the stability 
criterium $p{+}\Pi{\,>\,}0$ {\em anywhere} inside the freeze-out 
surface.     

\begin{figure}[t]
\centering{\includegraphics[width=\linewidth,clip=]{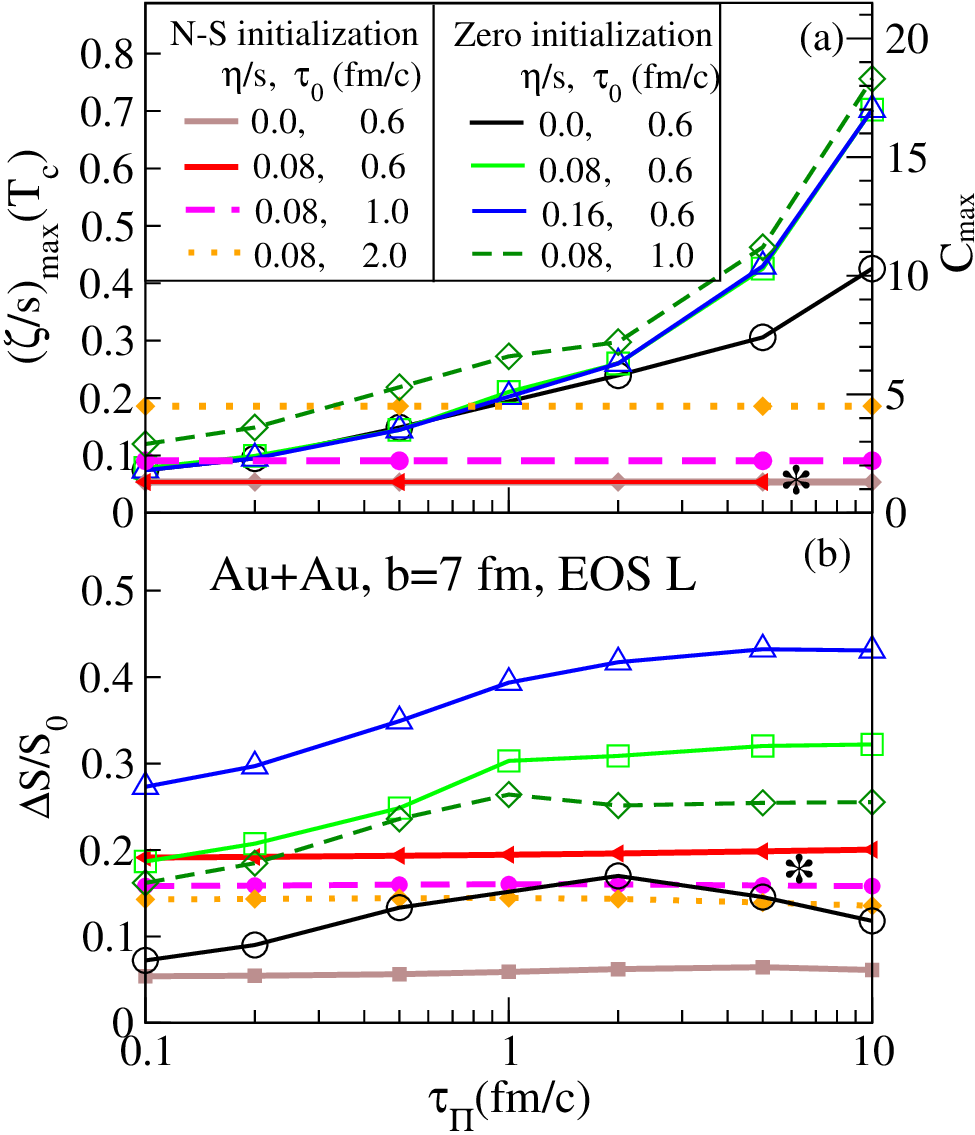}}
\caption{\label{F6} (Color online)
Upper limits for $\zeta/s$ (a) and viscous entropy production (b) as 
a function of bulk viscous relaxation time $\tau_{_\Pi}$, for zero and
Navier-Stokes initialization. The stars indicate the results for the 
temperature dependent relaxation time (\ref{eq1}) with Navier-Stokes
initial conditions, for $\tau_0\eq0.6$\,fm/$c$. For $\tau_{_\Pi}$ given 
by Eq.~(\ref{eq1}) and zero initial conditions, there is no upper limit 
for $\zeta/s$, {\it i.e.} the flud remains stable for all values of $C$.
}  
\end{figure}

Figure~\ref{F6} shows the upper limit  $(\zeta/s)_\mathrm{max}(T_c)$ 
(on the left vertical axis) and the corresponding maximal $C$-value
$C_\mathrm{max}$ (on the right vertical axis) as a function of the 
bulk viscous relaxation time $\tau_{_\Pi}$. We see that it depends 
strongly on the initialization. 

For Navier-Stokes (N-S) initial conditions $(\zeta/s)_\mathrm{max}(T_c)$ 
is insensitive to the relaxation time $\tau_{_\Pi}$. In this case the 
magnitude of the average bulk pressure $\Pi$ decreases more or less 
monotonically with time (see Fig.~\ref{F5}b). Vio\-lations of the positivity 
condition $p{+}\Pi{\,>\,}0$ thus always happen at the starting time 
$\tau_0$, at transverse positions where the matter is close to the phase 
transition. This leads to a $(\zeta/s)_\mathrm{max}(T_c)$ that is 
controlled by initial conditions and independent of the relaxation time. 
(This includes the temperature dependent relaxation time (\ref{eq1}) -- 
see the star in Fig.~\ref{F6}a.) Correspondingly, 
$(\zeta/s)_\mathrm{max}(T_c)$ does not depend on the value of $\eta/s$ 
when shear viscosity is included. When one starts the hydrodynamic 
evolution later, $(\zeta/s)_\mathrm{max}(T_c)$ increases with $\tau_0$. 
The dependence on $\tau_0$ arises from the strong dependence 
of the initial bulk pressure $\Pi\eq{-}\zeta\theta\eq{-}\zeta\partial{\cdot}u$ 
on $\tau_0$, through the expansion rate $\theta(\tau_0)\eq1/\tau_0$. 
This is illustrated by the solid red, dashed magenta and dotted orange 
lines in Fig.~\ref{F6}a: as one increases $\tau_0$ from 0.6 to 1 and 
2\,fm/$c$, the maximal $(\zeta/s)_\mathrm{max}(T_c)$ increases from 
0.05 to 0.09 and 0.18.

For zero initialization $\Pi(\tau_0)\eq0$ one finds a qualitatively
similar dependence of $(\zeta/s)_\mathrm{max}(T_c)$ on the starting
time $\tau_0$: The curves $C_\mathrm{max}(\tau_{_\Pi})$ move up 
as one increases $\tau_0$ from 0.6 to 1.0\,fm/$c$ (solid and dashed 
green lines). However, in contrast to the N-S initialization, the 
$(\zeta/s)_\mathrm{max}(T_c)$ curves now show a strong dependence on 
relaxation time $\tau_{_\Pi}$, rising monotonically with $\tau_{_\Pi}$. 
The reason is that it takes some time for the bulk pressure $\Pi$ to 
develop large enough magnitudes to violate the positivity condition 
$p{+}\Pi{\,>\,}0$; again this happens typically in regions where the 
matter is close to the phase transition. For larger relaxation times 
$\Pi$ moves away from its zero initial value more slowly, rendering the 
fluid more stable and resulting in a monotonic increase of 
$(\zeta/s)_\mathrm{max}(T_c)$ with $\tau_{_\Pi}$. For 
$\tau_{_\Pi}{\,<\,}1$\,fm/$c$ we find "universal"
$(\zeta/s)_\mathrm{max}(T_c)-\tau_{_\Pi}$ curves that do not depend on the
shear viscosity $\eta/s$ (solid black, green and blue curves), but
move upwards as we increase the starting time $\tau_0$. This is
because the violation of the positivity condition $p{+}\Pi{\,>\,}0$
then generally happens at early times $\tau{\,<\,}3$\,fm/$c$ when the
flow profiles are not yet significantly affected by shear viscous
effects. For the two viscous fluid lines with $\eta/s\eq0.08$ and 0.16
(solid blue and green lines) one sees that they continue to overlap even 
for $\tau_{_\Pi}{\,>\,}1$\,fm/$c$ after they have broken away from the 
$\eta/s\eq0$ line. In the ideal fluid ($\eta/s\eq0$) the phase transition 
generates large velocity gradients near the phase transition that generate 
locally large expansion rates, causing instability at lower values of 
$\zeta/s$. Shear viscosity smoothes out these large gradients, as discussed 
in Ref.~\cite{Song:2007ux}, allowing the fluid to evolve stably up to 
larger values of $\zeta/s$. Bulk viscosity $\zeta$ alone has no 
smoothing influence on sharp structures generated by a phase transition.
For a zero initial value for $\Pi$, shear viscosity thus helps crucially
in stabilizing the evolution of the viscous fluid against mechanical
instabilities caused by strongly negative bulk viscous pressure,
especially for large relaxation times $\tau_{_\Pi}$.

Very interesting is our finding that, for zero initial conditions, there
is no limit on $C_\mathrm{max}$ if one accounts for critical slowing down 
of the bulk pressure dynamics near $T_c$ via Eq.~(\ref{eq1}). In this 
case the bulk pressure, starting at zero, never grows sufficiently large
to threaten mechanical stability of the fluid, irrespective of how large 
the bulk viscosity becomes at $T_c$! As the peak value $(\zeta/s)(T_c)$
is increased, so is the time it takes $\Pi$ to evolve towards its 
Navier-Stokes value, and this never happens fast enough to violate
the stability condition $p{+}\Pi{\,>\,}0$.

Figure~\ref{F6}b shows the viscous entropy production for the maximally 
allowed bulk viscosities shown in Fig.~\ref{F6}a. Not surprisingly,
viscous entropy production increases with shear viscosity $\eta/s$ and 
decreases when hydrodynamics is started later, with correspondingly 
smaller initial expansion rates \cite{Song:2008si}. The dependence on 
$(\zeta/s)_\mathrm{max}(T_c)$ is non-monotonic, however. The reason is 
that the bulk viscous entropy production rate ${\sim\,}\Pi^2/(2\zeta)$
depends not only on how large $\zeta$ is but also on how close $\Pi$
is to its Navier-Stokes limit, and this in turn depends on $\tau_{_\Pi}$.    

\section{Towards extracting $\bm{\eta/s}$ from experimental data:
uncertainties introduced by bulk viscosity } 
\label{sec7}

\begin{figure}[t]
\includegraphics[width=\linewidth,clip=]{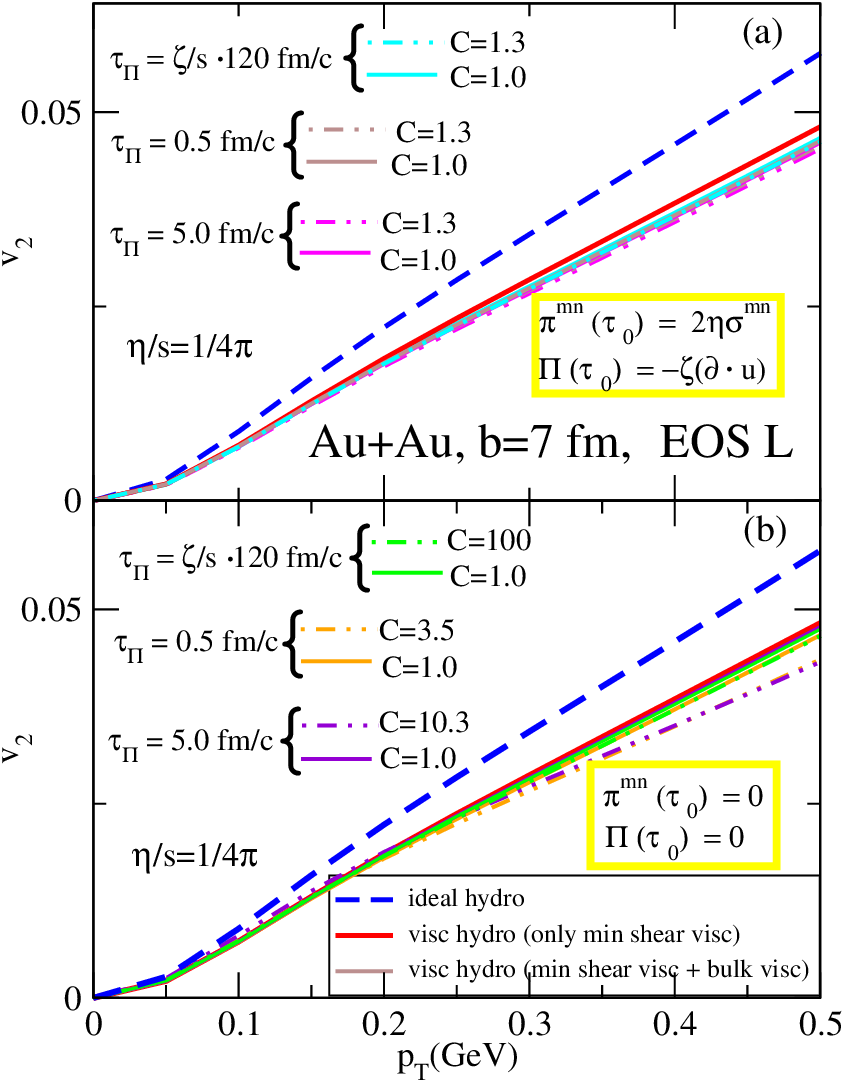}
\caption{\label{F7}
(Color online) $v_2(p_T)$ for directly emitted pions from
ideal and viscous hydrodynamics with Navier-Stokes (a) or zero (b)
initial conditions for the viscous pressures. Shown are results for 
minimal shear viscosity $\eta/s\eq1/4\pi$ and bulk viscosities ranging
from ``minimal'' ($C\eq1$) to the maximal values from Fig.~\ref{F6}a
that still allow for stable viscous evolution, for three choices of
the bulk viscous relaxation time $\tau_{_\Pi}$.
}
\end{figure}

\begin{table*}[htb]
\caption{\label{T1} Pion elliptic flow at $p_T\eq0.5$\,GeV/$c$ for 
$b\eq7$\,fm $200\,A$\,GeV Au+Au collisions from ideal and viscous 
hydrodynamics, with different choices of initial conditions, bulk
viscous relaxation times $\tau_{_\Pi}$, and bulk viscosities (parametrized 
by $C$). The last of the 3 columns in each initialization block gives
the viscous suppression of $v_2$ at $p_T\eq0.5$\,GeV/$c$ in terms of the 
percent deviation from the ideal fluid baseline ($\eq100\%$).} 
\begin{ruledtabular}
\begin{tabular}{|cc|ccc|ccc|}
  \multicolumn{2}{|c}{}
& \multicolumn{3}{c}{zero initialization}
& \multicolumn{3}{c|}{Navier-Stokes initialization}\\
\hline
\phantom{$\Big|_2$}
  $\quad\eta/s\quad\ $ & $\tau_{_\Pi}$\,(fm/$c$)$\quad$ 
& $C\ $ & $v_2(0.5\,\mathrm{GeV}/c)\,(\%)\ $ 
& $\frac{v_2}{v_\mathrm{2,ideal}}\,(\%)\quad$ 
& $C\ $ & $v_2(0.5\,\mathrm{GeV}/c)\,(\%)\ $ 
& $\frac{v_2}{v_\mathrm{2,ideal}}\,(\%)\quad$\\
\hline
\phantom{$\Big|$}
0 & -- & 0 & 5.755 & 100 & 0 & 5.755 & 100 \\
\hline
\phantom{$\Big|$}
0.08 & -- & 0 & 4.821 & 83.8 & 0 & 4.811 & 83.6 \\
\hline
\phantom{$\big|$}
0.08 & 0.5 & 1 & 4.668 & 81.1 & 1 & 4.627 & 80.4 \\
\phantom{$\big|$}
0.08 & 0.5 & 3.5 & 4.356 & 75.7 & 1.3 & 4.576 & 79.5 \\
\hline
\phantom{$\big|$}
0.08 & 5.0 & 1 & 4.770 & 82.9 & 1 & 4.601 & 79.9 \\
\phantom{$\big|$}
0.08 & 5.0 & 10.3 & 4.323 & 75.1 & 1.3 & 4.534 & 78.8 \\
\hline
\phantom{$\big|$}
0.08 & Eq.\,(\ref{eq1}) & 1 & 4.743 & 82.5 & 1 & 4.660 & 81.0 \\
\phantom{$\big|$}
0.08 & Eq.\,(\ref{eq1}) & 100 & 4.656 & 80.9 & 1.3 & 4.615 & 80.2 \\
\end{tabular}
\end{ruledtabular}
\end{table*}

Given the fact that bulk viscosity contributes to the viscous suppression
of elliptic flow (see Fig.~\ref{F5}a), and assuming that bulk and shear 
viscous effects cannot be separated by studying other experimental
observables, the question arises naturally how much of an irreducible 
uncertainty this will introduce into the extraction of the specific
shear viscosity $\eta/s$ from experimental elliptic flow measurements.  
More precisely, if the QGP should turn out to be a ``most perfect liquid''
with ``minimal'' shear viscosity $\eta/s\eq1/4\pi$, with what kind of
accuracy can we hope to verify this experimentally if bulk viscosity
is the only quantity beyond our theoretical and experimental control?

To answer this question, we used {\tt VISH2+1} to compute the differential
elliptic flow of directly emitted pions (without resonance decay 
contributions) for $200\,A$\,GeV Au+Au collisions at $b\eq7$\,fm, 
assuming the fireball medium to have constant specific shear viscosity
$\eta/s\eq1/4\pi$ but allowing the bulk viscosity $\zeta/s$ to vary over
the entire range allowed by the mechanical stability criterium
$p{+}\Pi{\,>\,}0$. In doing so we assumed a fixed shape of the temperature
dependence of $\zeta/s$ as shown in Fig.~\ref{F1} but let its normalization
vary between $C\eq1$ and $C_\mathrm{max}(\tau_{_\Pi})$ where the latter
is the largest value within the range of applicability of Israel-Stewart
viscous fluid dynamics, shown in Fig.~\ref{F6}a. We allowed for two
fixed values of 0.5 and 5\,fm/$c$ for the bulk viscous relaxation time
$\tau_{_\Pi}$ as well as for ``critical slowing down'' according to 
Eq.~(\ref{eq1}), and we studied both zero and Navier-Stokes initial 
values for the viscous pressure components. All calculations assume 
$\tau_0\eq0.6$\,fm/$c$ as starting time. The results are presented 
in Fig.~\ref{F7} and Table~\ref{T1}. 

Generically one observes that, even for minimal shear viscosity near the 
KSS bound, the shear viscous contribution to the elliptic flow suppression 
far exceeds the bulk viscous contribution. This is good news since it
means that the uncertainty introduced into the extraction of $\eta/s$ 
by theoretically poorly controlled bulk viscous effects remains limited 
and is, in fact, quite small, especially if the real fireballs created in 
heavy-ion collisions do not completely saturate the KSS bound. On a 
more quantitative level, one finds that for pions with typical 
transverse momentum $p_T\eq0.5$\,GeV/$c$ the elliptic flow is suppressed
by just over 16\% below the ideal fluid value if the expanding matter
has only shear, but no bulk viscosity, and that this suppression increases
to values between 17\% and 25\% if bulk viscosity is added. The largest
bulk viscous suppression is found for fixed relaxation times $\tau_{_\Pi}$ 
and zero initialization if the bulk viscosity is increased all the way
up to its upper allowed limit. In these cases the additional suppression 
can be as large as 50\% of the suppression found for the fluid with only
(minimal) shear viscosity. If one takes into account that the evolution
of the bulk viscous pressure slows down near $T_c$ where $\zeta/s$ is
largest, the additional bulk viscous suppression never exceeds 20\% of
the shear viscous elliptic flow suppression, with 10-15\% being a typical 
range (light blue and green curves in Fig.~\ref{F7}).

An important caveat is, however, that for Navier-Stokes initial conditions
the allowed maximal bulk viscosities are small, much below recent Lattice 
QCD estimates \cite{Meyer:2007dy}. If larger values are realized
by Nature, they invalidate the use of viscous hydrodynamics, at least 
at early times \cite{Dumitru:2007qr,Huovinen:2008te,Martinez:2009mf,%
Martinez:2009ry}. The problems in this case arise from the large bulk
viscosity in a thin layer near the transverse edge of the fireball
where the matter is close to $T_c$. It is only in this region that the
viscous hydrodynamic description breaks down. Since the problematic
factor, the scalar expansion rate $\theta$, decreases initially very 
rapidly, these initially unstable fluid regions move quickly back 
to mechancal stability. Since the momentum anisotropy does not develop 
instantaneously, we find it hard to believe that the existence of this 
unstable external layer has much influence on the evolution and final 
value of the elliptic flow, and one should get very similar results from
simulations in which the initial bulk viscous pressure $\Pi$ is restricted 
by hand to values below the threshold for violating the positivity
condition $p{+}\Pi{\,>\,}0$. If this is indeed the case, the results
presented in this Section show that bulk viscosity, even if theoretically 
not well controlled, will not introduce large uncertainties
into the extraction of $\eta/s$ from elliptic flow data. 

\section{Concluding remarks}
\label{sec8}

The present study shows that bulk viscosity, as long as it is small
enough that in expanding heavy-ion collision fireballs the negative 
bulk viscous pressure does not become larger than the thermodynamic 
pressure, affects the elliptic flow of the final hadrons much more 
weakly than does shear viscosity. So, as long as the expanding fireball 
can be described by viscous fluid dynamics, it is possible to extract 
its shear viscosity (even if it is as small as $\left.\frac{\eta}{s}
\right|_\mathrm{KSS}\eq\frac{1}{4\pi}$) with good accuracy from a 
comparison of viscous hydrodynamic simulations with experimental 
elliptic flow data. Accounting for the critical slowing down of 
viscous bulk pressure dynamics near $T_c$, we showed that any 
contamination from bulk viscosity $\zeta/s$ is ${<\,}20\%$ (for much 
of the parameter space it is even ${<\,}10\%$), and that its relative 
importance decreases further if $\eta/s$ is larger than the KSS bound.  

However, we also saw that the stability condition $p{+}\Pi{\,>\,}0$ 
is very restrictive and easily violated if the peak value of
$\zeta/s$ near $T_c$ reaches values close to those estimated from
Lattice QCD \cite{Meyer:2007dy} and from some strong coupling approaches
\cite{Buchel:2009hv}, {\em and} if the bulk viscous pressure $\Pi$
approaches its Navier-Stokes limit $\Pi\eq{-}\zeta\partial{\cdot}u$.
When this occurs (typically at early times when the scalar expansion 
rate is largest, in a thin layer around $T_c$ close to the transverse 
edge of the fireball), the viscous fluid dynamical description breaks
down. Our analysis shows that the phenomenon of ``critical slowing
down'' can play a crucial role in preventing this from happening.
Kinetic theory for weakly coupled systems \cite{Israel:1976tn,%
DeGroot:1980dk} and a recent ana\-ly\-sis by Buchel of strongly coupled
systems \cite{Buchel:2009hv} suggest that the same microscopic physics
(namely growing correlation lengths due to critical fluctuations) that 
generate a peak of $\zeta/s$ at $T_c$ also causes the relaxation time 
$\tau_{_\Pi}$ for the bulk viscous pressure to grow and possibly diverge 
at $T_c$ even while $\zeta/s$ itself remains finite. When using the model 
Eq.~(\ref{eq1}) for a temperature dependent $\tau_{_\Pi}$ inspired by 
these ideas we saw that, unless $\Pi$ is initialized at its Navier-Stokes 
limit, it never reaches it during the short time span of a heavy-ion 
collision in those fireball region where $\zeta/s$ peaks and $\Pi$ could 
thus become very large. This reduces the problem of applicability of 
viscous hydrodynamics at early times to a question of initial conditions 
for $\Pi$, especially in that thin transverse layer where (after local 
equilibrium is reached) the temperature happens to be close to $T_c$.

Determining these initial conditions (as opposed to guessing them as we 
have done here) requires a theoretical description of the early 
pre-equilibrium evolution and Landau-matching the corresponding
energy-momentum tensor to its viscous fluid dynamic form, Eq.~(\ref{eq2})
(in the spirit of Ref.~\cite{Martinez:2009ry} but generalized from 0+1
to 2+1 dimensions). At this point we lack the tools for doing this.
Let us, however, make a few comments in anticipation of completion of
that task. Consider a small fireball region that is just reaching local 
thermal equilibrium at a temperature close to $T_c$ and undergoing
self-similar boost-invariant longitudinal expansion while transverse 
expansion is negligible. Let us also assume that at this point in time
the bulk viscous pressure in the region is large and negative, leading
to negative effective total isotropic pressure and causing the fluid to 
be mechanically unstable. What will happen? The fluid will begin to
rupture, forming little voids, and if the region were to remain in a state 
of negative total pressure, it would eventually fragment. However, since 
the considered region is undergoing rapid expansion and cooling, it will 
quickly exit from its state of mechanical instability. Furthermore,
during the short period of instability the hydrodynamic growth of voids 
will be hampered by the large value of the relaxation time $\tau_{_\Pi}$.
By the time the considered region becomes mechanically stable again, we 
expect it to be riddled with small holes, but otherwise intact. The small 
voids formed during the period of instability will re-collapse by cavitation, 
and the region will quickly re-equilibrate due to the now much shorter 
relaxation time below $T_c$. No wholesale breakup of the fluid will occur, 
due to lack of time. Similar arguments hold later when the bulk of the 
matter in the center of the fireball passes through $T_c$, only that in 
this case the viscous bulk pressure may never grow large enough to generate 
mechanical instability, due to critical slowing down.

In summary, unlike the authors of Ref.~\cite{Rajagopal:2009yw}, we do not 
expect any dramatic macroscopic phenomena triggered by the transient 
mechanical instability arising from possibly large, but short-lived 
negative bulk pressures in fireball regions passing through the 
hadronization phase transition. For this reason we believe that a 
modified viscous hydrodynamic treatment, where one limits {\em by hand}
the growth of the viscous bulk pressure so that it always remains below the 
instability threshold \cite{Pratt:2007gj}, will not lead to impermissible 
distortions of the real (non-equilibrium) dynamics in the (small) 
space-time regions whose description lies outside the hydrodynamic 
domain. This is important for future viscous hydrodynamic studies of 
heavy-ion collisions with fluctuating and granular initial conditions 
\cite{Takahashi:2009na} which are more realistic than the smooth initial 
profiles presently used.

\vspace*{-3mm}
\acknowledgments{We gratefully acknowledge informative discussions with 
K. Dusling, P. Petreczky, and J. Randrup, and thank T.~Hirano and 
G.~Moore for constructive comments on the manuscript. U.H. is indebted 
to K.~Rajagopal for a very fruitful exchange of ideas which clarified 
much of the argument presented in the last Section of this paper. This 
work was supported by U.S.\ Department of Energy under contract 
DE-FG02-01ER41190.}



\begin{thebibliography}{99}
\bibitem{fn1}
  This argument assumes that the fluid is sufficiently weakly coupled 
  that a microscopic kinetic description in terms of quasiparticle 
  distribution functions makes sense. We do not know how to generalize 
  this argument to strongly coupled field theories that have no particle 
  interpretation. We suggest that the criterium $(|\Pi|{+}|\pi^{\mu\nu}|)/
  (e{+}p){\,\ll\,}1$ (where $|\pi^{\mu\nu}|$ is the largest viscous 
  pressure component) defines the domain of validity of ($2^\mathrm{nd}$ 
  order) viscous hydrodynamics, irrespective of the microscopic structure 
  of the fluid.
\bibitem{fn2}
  We note that in our calculations the bulk viscous pressure 
  $\Pi{\,\approx\,}0$ on the $T_\mathrm{dec}\eq130$\,MeV freeze-out 
  surface since our parametrization of the bulk viscosity gives 
  $\frac{\zeta}{s}(T_\mathrm{dec})\eq0$ (see Fig.~\ref{F1}). Accordingly,
  the bulk viscous pressure $\Pi$ does not contribute to the deviation
  $\delta\!f$ from local equilibrium on the freeze-out surface.
\end{thebibliography}
\end{document}